\documentclass[preprint2,times,tighten]{aastex6}

\pdfoutput=0 %for arXiv submission
\usepackage{amsmath,amstext}
\usepackage[all]{hypcap} %Figure refs go figures (breaks deluxetables; use \capstartfalse \capstarttrue to fix it)
\usepackage{placeins}
\usepackage{subfigure}

\hypersetup{
    colorlinks,
    linkcolor={red!50!black},
    citecolor={blue!50!black},
    urlcolor={blue!80!black}
}

\shorttitle{NGC 6791: Open Cluster with High-$\alpha$ Chemistry}
\shortauthors{Linden et al. 2016}

\begin{document}
\title{Timing the Evolution of the Galactic Disk with NGC 6791:\\ An Open Cluster with Peculiar High-$\alpha$ Chemistry as seen by APOGEE}
\author{Sean T. Linden\altaffilmark{1}, Matthew Pryal\altaffilmark{1}, Christian R. Hayes\altaffilmark{1}, Nicholas W. Troup\altaffilmark{1}, Steven R. Majewski\altaffilmark{1}, Brett H. Andrews\altaffilmark{2}, Timothy C. Beers\altaffilmark{3}, Ricardo Carrera\altaffilmark{4,5}, Katia Cunha\altaffilmark{6,7}, J. G. Fern\'andez-Trincado\altaffilmark{8}, Peter Frinchaboy\altaffilmark{9}, Doug Geisler\altaffilmark{10}, Richard R. Lane\altaffilmark{11}, Christian Nitschelm\altaffilmark{12}, Kaike Pan\altaffilmark{13}, Carlos Allende Prieto\altaffilmark{4}, Alexandre Roman-Lopes\altaffilmark{14}, Verne V. Smith\altaffilmark{7}, Jennifer Sobeck\altaffilmark{1}, Baitian Tang\altaffilmark{10}, Sandro Villanova\altaffilmark{10}, Gail Zasowski\altaffilmark{15}}
\altaffiltext{1}{Department of Astronomy, University of Virginia, Charlottesville, VA 22903, USA}
\altaffiltext{2}{PITT PACC, Department of Physics and Astronomy, University of Pittsburgh, Pittsburgh, PA 15260, USA}
\altaffiltext{3}{Department of Physics and JINA Center for the Evolution of the Elements, University of Notre Dame, Notre Dame, IN 46556, USA}
\altaffiltext{4}{Instituto de Astrof\'isica de Canarias, 38205 La Laguna, Tenerife, Spain}
\altaffiltext{5}{Departamento de Astrof\'isica, Universidad de La Laguna, E-38207 Tenerife, Spain}
\altaffiltext{6}{Observat\'orio Nacional, Rua General Jos\'e Cristino, 77, 20921-400 So Crist\'ovao, Rio de Janeiro, RJ, Brazil}
\altaffiltext{7}{National Optical Astronomy Observatory, 950 North Cherry Avenue, Tucson, AZ 85719, USA}
\altaffiltext{8}{Institut Utinam, CNRS UMR 6213, Universit\'e de Franche-Comt\'e, OSU THETA Franche-Comt\'e-Bourgogne, Observatoire de Be\-\\san\c{c}on, BP 1615, 25010 Besan\c{c}on Cedex, France}
\altaffiltext{9}{Department of Physics \& Astronomy, Texas Christian University, Fort Worth, TX 76129, USA}
\altaffiltext{10}{Department of Astronomy, Universidad de Concepci\'on, Chile}
\altaffiltext{11}{Unidad de Astronom\'ia, Universidad de Antofagasta, Avenida Angamos 601, Antofagasta 1270300, Chile}
\altaffiltext{12}{Instituto de Astrof\'isica, Pontificia Universidad Cat\'olica de Chile, Av. Vicuna Mackenna 4860, 782-0436 Macul, Santiago, Chile}
\altaffiltext{13}{Apache Point Observatory, P.O. Box 59, Sunspot, NM 88349, USA}
\altaffiltext{14}{Departamento de F\'isica, Facultad de Ciencias, Universidad de La Serena, Cisternas 1200, La Serena, Chile}
\altaffiltext{15}{Center for Astrophysical Sciences, Department of Physics and Astronomy, Johns Hopkins University, 3400 North Charles Street, Baltimore, MD 21218, USA}

\begin{abstract}
We utilize elemental-abundance information for Galactic red giant stars in five open clusters (NGC 7789, NGC 6819, M67, NGC 188, and NGC 6791) from the Apache Point Observatory Galactic Evolution Experiment (APOGEE) DR13 dataset to age-date the chemical evolution of the high- and low-$\alpha$ element sequences of the Milky Way. Key to this time-stamping is the cluster NGC 6791, whose stellar members have mean abundances that place it in the high-$\alpha$, high-[Fe/H] region of the [$\alpha$/Fe]-[Fe/H] plane. Based on the cluster's age ($\sim 8$ Gyr), Galactocentric radius, and height above the Galactic plane, as well as comparable chemistry reported for APOGEE stars in Baade's Window, we suggest that the two most likely origins for NGC 6791 are as an original part of the thick-disk, or as a former member of the Galactic bulge. Moreover, because NGC 6791 lies at the \textit{high metallicity end} ([Fe/H] $\sim 0.4$) of the high-$\alpha$ sequence, the age of NGC 6791 places a limit on the \textit{youngest age} of stars in the high-metallicity, high-$\alpha$ sequence for the cluster's parent population (i.e., either the bulge or the disk). In a similar way, we can also use the age and chemistry of NGC 188 to set a limit of $\sim 7$ Gyr on the \textit{oldest age} of the low-$\alpha$ sequence of the Milky Way. Therefore, NGC 6791 and NGC 188 are potentially a pair of star clusters that bracket both the timing and the duration of an important transition point in the chemical history of the Milky Way.

%The age of NGC 6791 also well-matches the transition time adopted in models attempting to explain the evolution from a high- to a low-$\alpha$ chemistry in the Milky Way using gas infall.
\end{abstract}

\keywords{Open Clusters and Associations: individual (NGC 6791), Stars: Abundances, Galaxy: Disk, Galaxy: Evolution, Galaxy: Stellar Content}
\maketitle

\section{Introduction}
The variation of stellar chemical abundances within galaxies contains information about how galaxies are assembled, and tracks processes such as gas accretion, star formation, and stellar migration (e.g., \citealt{Matteucci&Francois 1989,Gilmore et al. 1989,Prantzos&Aubert 1995,Chiappini et al. 2001,Hayden et al. 2015}). However, the origin and chemical evolution of the Milky Way (MW) disk system has remained a much debated topic over the past few decades, since the discovery of the ``thick-disk'' as a major structural component of the Galaxy 
(e.g., \citealt{Yoshii 1982,Gilmore&Reid 1983,Gilmore et al. 1985,
Majewski 1993,Bovy et al. 2012a,Bovy et al. 2012b,Robin et al. 2014,Masseron&Gilmore 2015}), 
though these disk-like stars with extreme metallicity and kinematics have been recognized for at least half a century (as, e.g., the ``intermediate Population II'' at the 1957 Vatican conference; \citealt{O'Connell 1957}).

More recently, attention to these issues has been fueled by new data from large, detailed chemical evolution surveys of MW field stars. In the solar neighborhood, the distribution of stars in the [$\alpha$/Fe] (where $\alpha$, represents elements such as O, Mg, Si, S, Ca, or Ti) to metallicity (e.g., [Fe/H]) plane displays two distinct sequences, the ``high-$\alpha$'' and ``low-$\alpha$'' sequences \citep{Garcia Perez et al. 2016}.

The high-$\alpha$ sequence extends from metal-poor stars with enhanced 
[$\alpha$/Fe] ratios, to stars of roughly solar abundances (e.g. \citealt{Bensby et al. 2007}). The shape of this sequence is maintained more or less ubiquitously across the Galaxy
\citep{Nidever et al. 2014,Hayden et al. 2015}. These high-$\alpha$ stars are believed to be associated with the thick disk of the MW due to their dynamical phase space 
%SRM5: I'm not sure we need to put in the explicit definition of phase space since it is a well-defined thing that everyone should know?
%($x, y, z, U, V, W$) 
distribution \citep{Gratton et al. 1996,Fuhrmann 1998,Prochaska et al. 2000,Bensby et al. 2003,Reddy et al. 2006,Adibekyan et al. 2011,Bensby et al. 2014, Hayden et al. 2015}. Scenarios for the formation of the thick disk have been influenced by its apparent age, location in phase space, and overall abundances relative to the thin disk --- these include scenarios that rely on external satellite accretion \citep{Quinn et al. 1993,Brook et al. 2004}, or on secular evolution of the disk through dissipation, collapse, heating, and/or radial migration \citep{Larson 1976,Gilmore 1984,Sellwood&Binney 2002}. Now that the thick disk is commonly associated with the high-$\alpha$ sequence seen in chemical-abundance studies, additional models have been inspired by taking into account its detailed chemistry. For example, \citet{Nidever et al. 2014} found that the high-$\alpha$ sequence can be explained with a leaky-box galactic chemical evolution (GCE) model that uses a nearly constant star formation efficiency (SFE) and a well-mixed, turbulent, and molecular-dominated interstellar medium (ISM), consistent with observations from \citet{Leroy et al. 2008}. 

In these chemical-evolution models of the disk, the origin and evolution of the low-$\alpha$ sequence is less well understood. In contrast with the high-$\alpha$ sequence, the location of the peak density of the low-$\alpha$ distribution in the [$\alpha$/Fe]-[Fe/H] plane varies with both Galactocentric radius and height above the Galactic plane \citep{Hayden et al. 2015, Nidever et al. 2014}. Despite this, the overall bi-modality of the disk [$\alpha$/Fe] - [Fe/H] distribution suggests the formation of the thin disk as a distinct, second population of stars, which has been proposed to have originated from a sudden infall of pristine gas $\sim 8$ Gyr ago, echoing previous suggestions from \citet{Chiappini et al. 1997}. However, \citet{Haywood et al. 2013} and \citet{Feuillet et al. 2016} used samples of field stars to show that this scenario may be difficult to reconcile with the overlap seen in the ages of outer-disk stars found in the metal-rich end of the high-$\alpha$ sequence and the metal-poor end of the low-$\alpha$ sequence. If born out, this age overlap means either that the transition was not so rapid in the case of a synchronized, Galaxy-wide chemical evolution, or that the transition might have been rapid but the timing of it varied as a function of position in the Galaxy. Such age discrepancies remain one of the key issues facing models that attempt to explain the two $\alpha$-element sequences in a coherent picture of galaxy-wide MW star formation.

Despite the enormous progress that has been made in exploring the detailed chemistry of the disk with increasingly larger samples (e.g., LAMOST - \citealt{Xiang et al. 2017}; GALAH - \citealt{De Silva et al. 2015}; Gaia-ESO - \citealt{Jacobson et al. 2016,Smiljanic et al. 2016}), field stars nevertheless suffer from various disadvantages that complicate their use in ascertaining the global properties of the disk, including (a) uncertainties in extinction, intrinsic luminosity, and therefore the distances of individual stars, (b) susceptibility to radial migration, which obscures stellar birth locations and presumably blurs disk properties tracked by location, kinematics, metallicity or age, and (c) large systematic and random uncertainties in deriving ages for individual stars \citep{Clem et al. 2004,Masana et al. 2006} --- although there is new promise in resolving this latter problem using the techniques of asteroseismology, gyrochronology (e.g., \citealt{Angus et al. 2015,Barnes et al. 2016}), and even detailed stellar chemistries for red giant stars (e.g., \citet{Ness et al. 2016}).

On the other hand, open clusters have long been used as a key Galactic tracer for probing chemical and age distributions within the MW disk (e.g., \citealt{Janes 1979,Panagia&Tosi 1980,Carraro&Chiosi 1994,Janes&Phelps 1994,Friel 1995, Cunha et al. 2016}). This is because open clusters behave like simple stellar populations, which enables measurement of reliably-determined ages, distances, metallicities, and, by combining spectroscopic data over many member stars, chemical-abundance patterns. Open clusters are especially  valuable as a population tracer probing low Galactic latitudes. Moreover, while clusters can be dynamically heated over time, young open clusters are less likely than individual field stars to have moved significantly from their birth orbits \citep{Friel 1989,Wu et al. 2009}.

In this paper, we exploit some of these advantages, and begin to build a sample of open clusters with reliable ages, distances, and especially, elemental abundances, to study the chemical evolution of the MW thick+thin disk system. We leverage the high-quality abundances of cluster red giant stars obtained as part of the calibration cluster data set in the Apache Point Observatory Galactic Evolution Experiment (APOGEE; \citealt{Majewski et al. 2015}). We focus in particular on the location of these open clusters among the $\alpha$-element sequences discussed above, and use the ages of these clusters to place time limits on the key phases of disk formation in the scenarios proposed by \citet{Haywood et al. 2013} and \citet{Nidever et al. 2014}. An additional outcome of this analysis is to add a chemically-based challenge to the traditional notion that objects classified as open clusters are exclusively denizens of the thin disk.

\section{Data}
The Apache Point Observatory Galactic Evolution Experiment (APOGEE; \citealt{Majewski et al. 2015}) is a near-infrared, high-resolution spectroscopic survey of predominantly red giant stars across the MW, included as part of the third and fourth Sloan Digital Sky Survey (\citealt{Gunn et al. 2006}; 
%SDSS-III: 
\citealt{Eisenstein et al. 2011}; 
%SDSS-IV: 
\citealt{Blanton et al. 2017}). 
APOGEE's $H$-band spectra, at a resolution resolving power of $R\sim$22,500 allow for the measurement of the chemical abundances of more than 15 chemical elements (\citealt{Holtzman et al. 2015}, \citealt{Shetrone et al. 2015}), some not accessible to from optical spectroscopy. Though taken as part of the APOGEE observations of some 146,000 stars in SDSS-III from 2011-2014, the data used in 
\begin{deluxetable*}{cccccccccc}
\center
\tabletypesize{\footnotesize}
\tablecolumns{10}
\tablewidth{0pt}
\tablecaption{Adopted Representative Properties for the Open Cluster Sample}
\tablehead{
\colhead{Name} & \colhead{RA} & \colhead{Dec} & \colhead{Age (Gyr)\tablenotemark{a}} & \colhead{$[Fe/H]$\tablenotemark{b}} & \colhead{$E(B-V)$\tablenotemark{c}} & \colhead{$N_{stars}$\tablenotemark{d}} & \colhead{Z (kpc)\tablenotemark{e}} & \colhead{$R_{GC}$ (kpc)\tablenotemark{e}} & \colhead{Tidal Radius\tablenotemark{f}}}
\startdata
NGC 7789 & $23^{\rm h} 57^{\rm m} 24.0^{\rm s}$ & $+56\arcdeg 42' 30"$ & 1.4 & -0.01 & 0.28 & 16 & -0.16 & 8.92 & 46' \\
NGC 6819 & $19^{\rm h} 41^{\rm m} 18.1^{\rm s}$ & $+40\arcdeg 11' 12"$ & 2.5 & 0.14 & 0.14 & 41 & 0.348 & 7.69 & 38' \\
M67 & $08^{\rm h} 51^{\rm m} 20.1^{\rm s}$ & $+11\arcdeg 48' 43"$ & 4.0 & -0.01 & 0.04 & 32 & 0.48 & 8.64 & 36' \\
NGC 188 & $00^{\rm h} 47^{\rm m} 27.5^{\rm s}$ & $+85\arcdeg 16' 11"$ & 7.0 & 0.03 & 0.09 & 2 & 0.78 & 9.17 & 45' \\
NGC 6791 & $19^{\rm h} 20^{\rm m} 53.2^{\rm s}$ & $+37\arcdeg 46' 19"$ & 8.3 & 0.38 & 0.14 & 5 & 1.107 & 8.09 & 23' \\
\enddata
%\tablenotetext{}{}
\tablenotetext{a}{Estimates for the age of each cluster: NGC 7789's age is taken from \citet{Wu et al. 2007}. M67's age is taken from the average of \citet{Schiavon et al. 2004}, \citet{Salaris et al. 2004}, and \citet{Barnes et al. 2016}. NGC 6819's age is taken from \citet{Kalirai et al. 2001}. NGC 188's age is taken from \citet{Bonatto et al. 2005}. NGC 6791's age is taken from \citet{Brogaard et al. 2011}.}
\tablenotetext{b}{Metallicity of each cluster taken from \citet{Dias et al. 2002}, \citet{Frinchaboy et al. 2013}, and \citet{Meszaros et al. 2013}.}
\tablenotetext{c}{Estimates for extinction along the line of sight to the cluster taken from the NASA Extragalactic Database (NED).}
\tablenotetext{d}{Number of cluster stars we identify in the APOGEE DR13 dataset.}
\tablenotetext{e}{Estimates for the Galactocentric radius and distance from the Galactic plane taken from \citet{Wu et al. 2009}.}
\tablenotetext{f}{Tidal radii, in arcminutes, taken from \citet{Piskunov et al. 2008} for NGC 7789 and M67, \citet{Yang et al. 2013} for NGC 6819, \citet{Wang et al. 2015} for NGC 188, and \citet{Platais et al. 2011} for NGC 6791.}
\end{deluxetable*}
\noindent
this paper come from spectra that have been re-reduced, re-calibrated, and made publicly available as part of the recent DR13 data release \citep{Albareti et al. 2016} of SDSS-IV. The DR13 reduction of the APOGEE data allows examination of the $\alpha$-sequence with more precise detail enabled by the most recent version of the
APOGEE Stellar Parameters and Chemical Abundances Pipeline (ASPCAP; \citealt{Garcia Perez et al. 2016}).

For our study, five open clusters were selected from the list of APOGEE calibration open clusters in \citet{Meszaros et al. 2013}. We use these particular clusters because they have been critical to calibrating the APOGEE abundance scale, due to significant pre-existing literature studies that have produced reliable age, metallicity, and extinction estimates. In addition, these five clusters have among the best APOGEE coverage and span the range in age and metallicity for disk clusters used as calibration clusters. The calibration stars targeted by APOGEE in these clusters are flagged in the DR13 database by the $APOGEE\_CALIB\_CLUSTER$ flag as $APOGEE\_TARGET2 = 10$ \citep{Zasowski et al. 2013}. The adopted age, metallicity, position, and extinction for each cluster in Table 1 are chosen to be representative of the full distribution of values in the literature.

%NGC 6791, in particular, stands out in the sample as the oldest open cluster studied. NGC 6791 also has the largest distance above the plane ($Z_{GC}$), [M/H], and [$\alpha$/M] in the sample. 
%\pagebreak

\begin{figure*}
\centering
\includegraphics[scale=0.35]{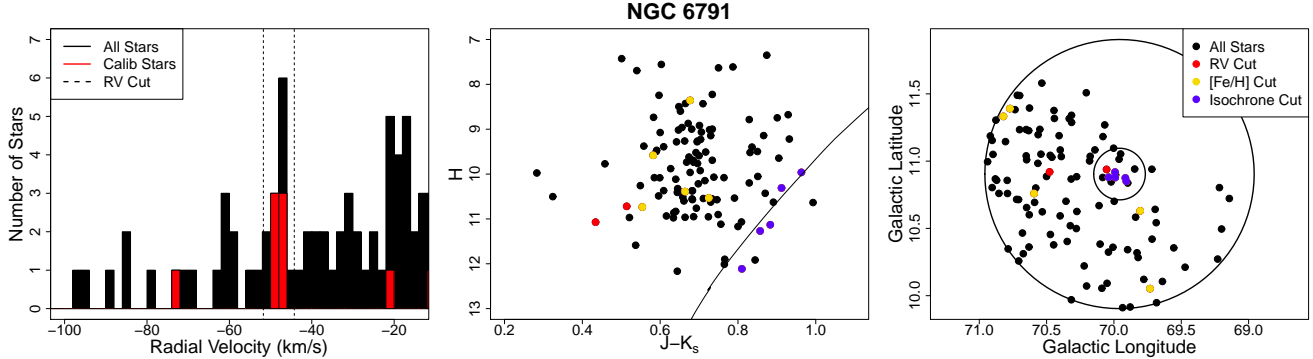}
\caption{Radial velocity histogram, color-magnitude diagram, and spatial distribution of stars in NGC 6791. In the left panel, the red bars represent stars flagged as calibration cluster stars and the dotted lines represent our 1.5$\sigma$ cut on the radial velocity histogram peak centered on -48 $\rm{km\ s}^{-1}$ with a standard deviation of 2.5 $\rm{km\ s}^{-1}$. In the middle and right panels, all colored points are stars flagged as potential cluster members based on radial velocity cuts alone, with yellow signifying stars that additionally passed the [Fe/H] cut, and blue signifying our final sample of identified cluster members after application of the additional isochrone and spatial cuts. The isochrone was created with [Fe/H]$=0.38$, $[\alpha$/Fe]$=0.2$, and an age of 8.3 Gyr using the Dartmouth Stellar Evolution Isochrone generator \citep{Dotter et al. 2007}. The inner circle is half the published tidal radius of the cluster and the outer circle has a radius of 1 degree, centered on the cluster central coordinates.}
\label{fig:n6791_rv_cmd}
\end{figure*}

To study individual clusters, we must first identify which stars in the observed APOGEE cluster fields are cluster members. Starting from the APOGEE calibration sample, we sought to increase the number of member stars from each cluster to include in our analysis based on a star's spatial location, radial velocity, and [Fe/H] relative to the centroids for the pre-verified calibration cluster members in the DR13 database. We also consider the star's position relative to the best-fit isochrone to the apparent cluster locus in the color-magnitude diagram (CMD). Because of the large fields-of-view of the APOGEE fields, we limit our selection to stars within $1^{\circ}$ of the established cluster centers as given in the literature and the SIMBAD database. We then created radial velocity histograms of the remaining cluster candidates in 2 $\rm{km\ s}^{-1}$ bins, as shown in Figure \ref{fig:n6791_rv_cmd} (left panel), where the red line signifies stars tagged with $APOGEE\_CALIB\_CLUSTER$ in $APOGEE\_TARGTET2$ as being a known calibration cluster member. We required cluster members to be within $1.5\sigma$ from best-fit Gaussians to the peak of each cluster's radial velocity histogram, as shown by the dotted lines in the left panel of Figure \ref{fig:n6791_rv_cmd}. For NGC 6791, the peak of the distribution fell at -48 $\rm{km\ s}^{-1}$ with a standard deviation of 2.5 $\rm{km\ s}^{-1}$.  The initial $1^{\circ}$ cluster radius selection allowed radial velocity peaks of the potential cluster members to stand out significantly amongst the field stars, as seen, for example, in Figure \ref{fig:n6791_rv_cmd}. Membership identification plots for the four other clusters can be seen in Figure \ref{app:member_plots}.
%SRM5: I wonder if it violates ApJ/AJ policy to have appendices that have no corresponding text -- someone should check.  Is there any reason to do these  extra plots as appendices, rather than just include them in the main text?  

% SRM5: But please don't do the usual thing of cutting and pasting the same caption for each figure?  Either do "same as Figure 1, but or NGC XXXX and with an isochrone of age x.x Gyr.", or change up the form of the caption for each one.  You could also just make all four figures one combined figure -- you could fit them all on one page with one caption.  There are titles on each one  already to keep them clear.  Ditto for Figures in Appendix B.  By the way, why is the numbering of the figures using letters and not starting with A1 and B1?
%MP5: I'm not sure why the numbering is the way it is but that is just what the aastex automatically did, I didn't force it to do that. I'll combine the plots into a single figure with caption giving additional info for each. I could not find anywhere stating that text must appear in appendices but I could always just throw in text so we don't have to worry about it.

%SRM5: Can the left panels of these figures be box histograms?  I have trouble reading these low statistics "point to point" versions of  histograms.

%SRM5: By the way, Table 1 is missing a cricital parameter -- the 
% adopted distance and/or distance modulus. THis needs to be given because it affects the plotted isochrone in the Figure and the derived RGC and ZGC values in the table. Also, what did you adopt for the conversion between E(B-v) and E(J-K), and also for A_K as a function of E(B-V)?  You have to give this as well.

Additionally, we required that cluster candidates be within 0.3 dex of the average [Fe/H] for the pre-verified calibration cluster members, as shown in Figure \ref{fig:n6791_rv_cmd} (middle and right panels). The $(J-K_{s}, H)$ CMD of all stars that are within $1^{\circ}$ of the cluster center are plotted for the NGC 6791 example, with all colored points representing those stars that also pass the radial velocity membership criterion. Gold and blue points indicate stars that pass the radial velocity selection and have measured metallicities within 0.3 dex of the mean abundance for the calibration stars in each cluster. Our next step was to make cluster cuts based on position relative to a roughly best-fit isochrone in the CMD.

For NGC 6791, the isochrone shown in the middle panel of Figure \ref{fig:n6791_rv_cmd} was created using the Dartmouth Stellar Evolution Isochrone generator \citep{Dotter et al. 2007} with [Fe/H]=0.38, [$\alpha$/H]=0.2 and an age of 8.3 Gyr. From \citet{Cardelli et al. 1989} we obtain the adopted $J-K_s$ extinction relation of $E(J-K_s) = 0.52\ E(B-V)$. With this exploration of the CMD distribution, we are able to remove stars that clearly were separate from other cluster members.
 
Our final step in the cluster member identification procedure is to impose another, more strict spatial limit, where needed, by requiring cluster candidates to be within half the known tidal radius of the cluster center, to further improve the likelihood of securing certain cluster members. The assumed tidal radii and their sources for each cluster are included in Table 1. This last criterion was imposed to ensure that the stars eventually chosen to be cluster members are likely to actually be under the influence of the cluster potential. In the end, the adopted combined selection criteria were rather conservative, to obtain samples as pure as possible to represent the mean chemistry of the cluster, at the price of likely cutting some cluster members that may be in the APOGEE catalog.

The result of this multistep selection process is shown for the NGC 6791 example in Figure \ref{fig:n6791_rv_cmd} (right panel), where the blue symbols denote the final cluster member identifications that pass all of our adopted criteria. For our analysis, the final sample consists of 96 stars from the five open clusters (Table 1), specifically: 16 from NGC 7789, 32 from M67, 41 from NGC 6819, 2 from NGC 188, and 5 from NGC 6791. A list of all notable parameters for individual stars identified in each cluster is provided in Table 2. The reference solar abundance values adopted for our analysis are given in \citet{Asplund et al. 2005}.

We note that three of the open clusters (NGC 6791, M67, and NGC 6819) in our study have been shown independently to be internally very chemically homogeneous, especially for the elements iron, sodium, and oxygen, at the dispersion level of $\sim 0.05-0.07$ dex for NGC 6791 \citep{Cunha et al. 2015} and $\sim 0.02-0.05$ dex for M67 and NGC 6819 respectively \citep{Bovy 2016}. Thus the abundance scatter we see in these clusters is likely dominated by the measurement uncertainty from the DR13 analysis of the APOGEE spectra.
%SRM5: Can you cite "to within 0.0x dex"?  And does this statement apply across multiple chemical elements or certain ones?  THis is helpful to support and clarify the meaning of your following statement.

We also note that our star selection for NGC 6791 represents only 5 of the 11 stars used in \citet{Cunha et al. 2015}. Of the eleven stars used in their independent (i.e., non-ASPCAP) analysis we exclude four automatically due to the stars being cooler than $T_{\rm eff} = 4000$ K, where the ASPCAP abundances are poorly constrained. The other two were eliminated from our analysis due to our conservative signal-to-noise criteria ($S/N \geq 100$), but, in fact, these stars have ASPCAP-derived abundances in line with the mean values of the five NGC 6791 stars included in our analysis, and so would not have affected our results had we modified our selection criteria to include them. When comparing mean abundances of [O/Fe] and [Na/Fe] for the matching NGC 6791 stars in our sample we find good agreement with the \citet{Cunha et al. 2015} results. Additionally, the individual oxygen and sodium abundances all agree within the combined 1$\sigma$ uncertainties of each analysis of the database.

\begin{figure*}
\centering
\includegraphics[scale=0.6]{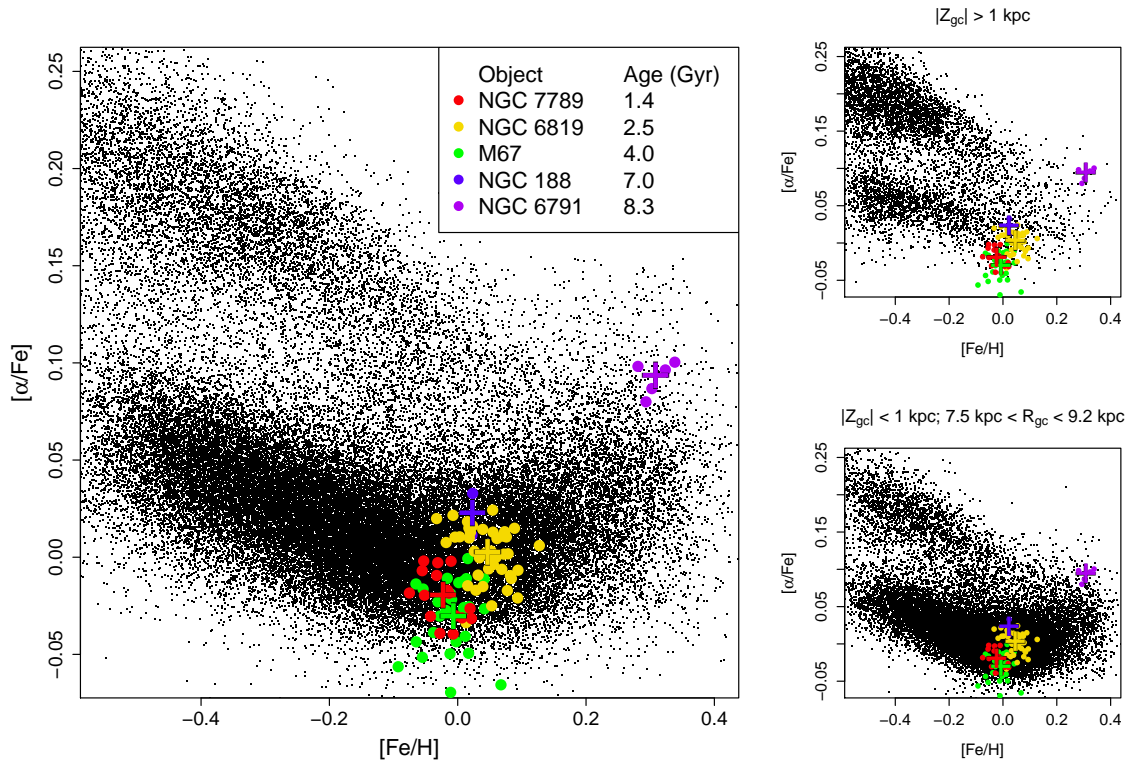}
\caption{The chemical evolution of the Milky Way as traced by the abundance of total $\alpha$-elements versus overall metallicity as  traced by [Fe/H]. The colored points represent stars identified as members in the five open clusters in our study while the black points are additional members of the APOGEE DR13 dataset with all stars (left), all stars with $|Z_{gc}| > 1$ kpc (upper right), and all stars with $|Z_{gc}| < 1$ kpc and $7.5$ kpc $< R_{gc} < 9.2$ kpc (lower right). The colored cross symbols signify the weighted mean abundances for each individual cluster. The median error in [$\alpha$/Fe] is 0.02 dex and in [Fe/H] is 0.03 dex for all stars.}
\label{fig:alpha_seq}
\end{figure*}

\section{Results}

Figure \ref{fig:alpha_seq} shows the [$\alpha$/Fe]-[Fe/H] distribution for the identified cluster members plotted against MW field stars in the DR13 dataset. Measured properties of individual cluster member stars for all five clusters examined can be found in Table 2.%\ref{tab:table}. 

Because they may have potentially larger errors in derived properties from their APOGEE combined spectra, we cut from the total field star sample any stars with: (1) an absolute error in the measured radial velocity of larger than $0.2$ km/s, (2) a visit-to-visit velocity scatter of $\geq 1$ $\rm{km\ s}^{-1}$, (3) a $T_{\rm eff} \leq 4000$ K, and (4) a $S/N \leq 100$. We additionally split the field star sample into a low-$Z_{GC}$ ($Z_{GC} \leq 1$ kpc) solar-circle ($R_{\odot} \sim 8$ kpc) dataset and a high-$Z_{GC}$ ($Z_{GC} \ge 1$ kpc) dataset. Galactocentric radii were calculated using distance data for APOGEE stars from \citet{Santiago et al. 2016}, and adopting a solar Galactocentric distance of 8 kpc. As expected, the high-$Z_{GC}$ cut does an efficient job of removing the majority of the low-$\alpha$ sequence primarily associated with the thin-disk of the MW.

\begin{figure}
\centering
\includegraphics[scale=0.43]{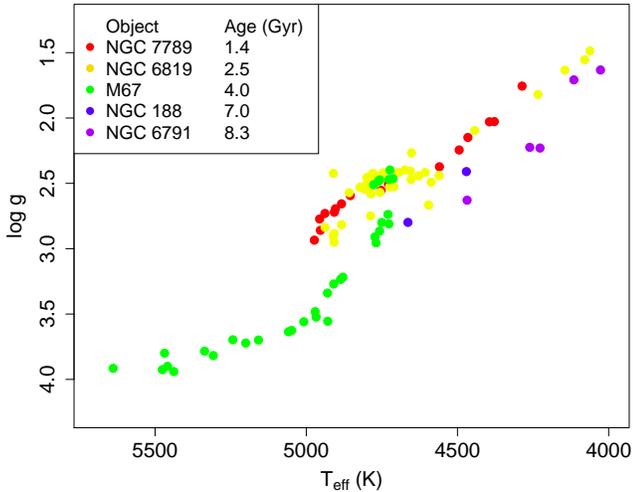}
\caption{Surface gravity ($\log g$) versus effective temperature ($T_{\rm eff}$) for stars identified in each cluster analyzed.}
\label{fig:T_logg}
\end{figure}

We additionally checked the stellar characteristics of identified cluster members to ensure there was no inherent biasing in our chemical abundance analysis by the actual stars identified. Figure \ref{fig:T_logg} plots surface gravity versus effective temperature for each of the cluster members. The clear spread of stars within each cluster for these properties shows we are sampling a relatively large distribution of surface gravity and temperature across the entire sample and within each cluster itself. If our abundances were being systematically biased by poorly constrained stellar characteristics as a result of these large distributions, we would expect to see an equally large spread in the mean abundance values for each of the clusters. This effect is not seen in any of the five clusters studied here. In particular, NGC 6791, which, as we shall show exhibits quite distinct chemical abundance properties (Figure \ref{fig:alpha_seq}), is sampled in log $g$ - $\rm{T_{eff}}$ space quite similarly to NGC 7789, NGC 6819 and NGC 188, and has been sampled in at least a partly overlapping log $g$ range with M67 (Figure \ref{fig:T_logg}). Moreover, a one-to-one comparison of the stars in NGC 6791 manually analyzed in \citet{Cunha et al. 2015} shows agreement to within 0.1 dex for the DR13 ASPCAP values, which re-affirms the accuracy and consistency of this pipeline.

%MP note: The referee states that giants at the high metallicity for example in 6791 have issues analyzing but I found that the errors associated with element abundances was on average lowest in 6791...

The four youngest open clusters studied here lie along the densest region of the low-$\alpha$ sequence that is associated with the thin disk (and in the part of that sequence dominated by stars near the solar circle), and are clearly separated from the high-$Z_{GC}$ track (Figure \ref{fig:alpha_seq}). As expected, these clusters are clearly part of the thin disk, and as a group provide some information on the age spread of stars in the low-$\alpha$ sequence.

However the $\lesssim 7$ Gyr age distribution of the five clusters places only a lower limit on the maximum age for the low-$\alpha$ sequence, since these objects lie roughly in the middle of the [Fe/H] range of that sequence. That said, if one compares the relative positions of the four younger clusters within the low-$\alpha$ sequence, there is an unexpected (though slight, and therefore tentative) correlation of the clusters' mean [$\alpha$/Fe]-[Fe/H] position with age, in the sense that the most metal-rich thin-disk cluster (NGC 188) is the oldest and the most metal-poor thin-disk cluster (NGC 7789) is the youngest. This trend, derived from a small sample of four clusters, may be entirely spurious, but, if not, could perhaps reflect a poorly mixed ISM during the formation of these clusters.

\begin{figure*}
\centering
\includegraphics[scale=0.58]{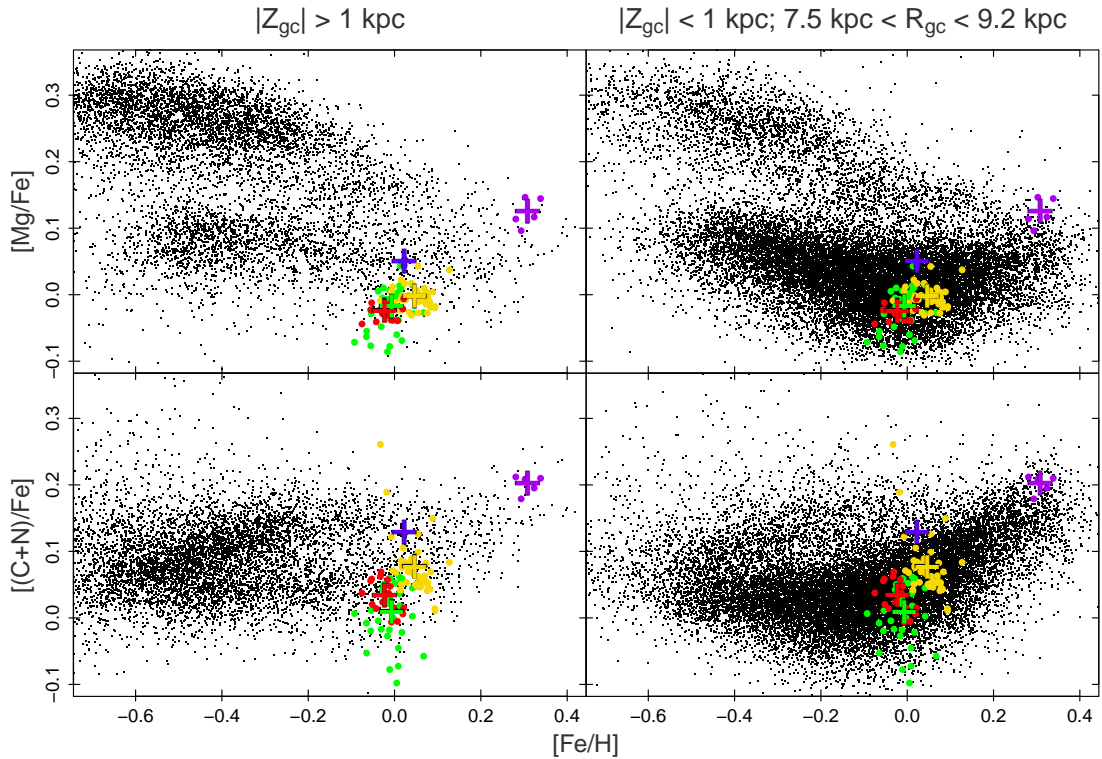}
\caption{The chemical evolution of the Milky Way as traced by the abundance of the $\alpha$-elements magnesium and total $C+N$ abundance versus metallicity, [Fe/H]. The colored points represent stars identified as members in the five open clusters in our study and the black points are additional members of the DR13 dataset with stars having $|Z_{gc}| > 1$ kpc (left panels), and $|Z_{gc}| < 1$ kpc but orbiting near the solar circle (right panels). The colors for each cluster match those adopted in Figure \ref{fig:alpha_seq}.}
\label{fig:mg_ca}
\end{figure*}

\begin{figure*}[t!]
\label{app:member_plots}
    \centering
    \begin{subfigure}%{0.35\textwidth}
        \centering
		\includegraphics[scale=0.34]{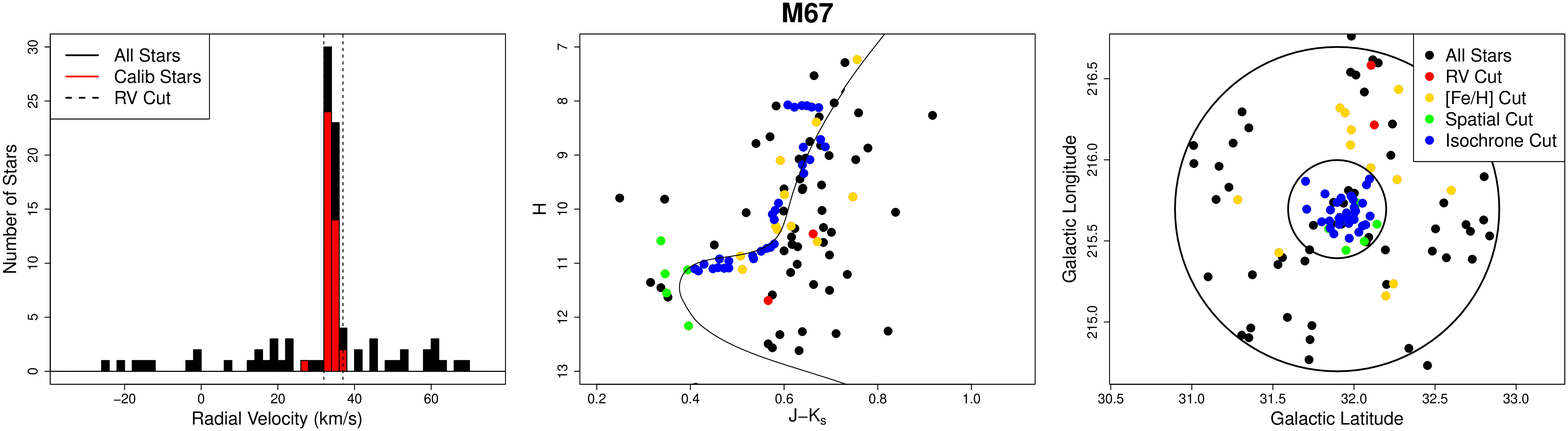}
		\label{fig:m67_rv_cmd}
    \end{subfigure}%
    ~ 
    \begin{subfigure}%{0.5\textwidth}
        \centering
		\includegraphics[scale=0.34]{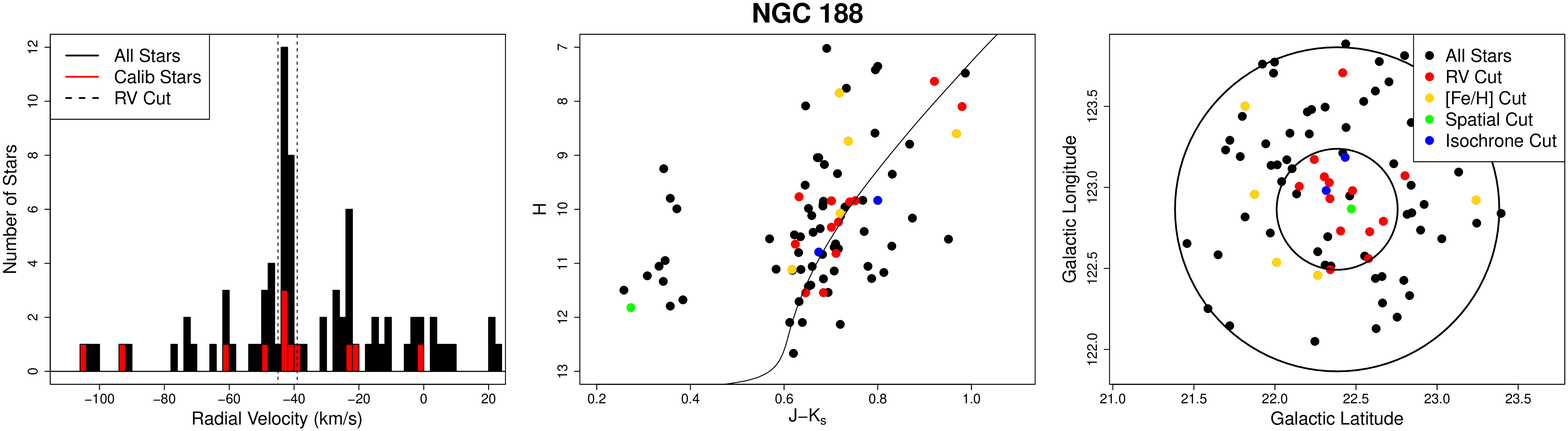}
		\label{fig:n188_rv_cmd}
    \end{subfigure}
    ~
    \begin{subfigure}%{0.35\textwidth}
        \centering
		\includegraphics[scale=0.34]{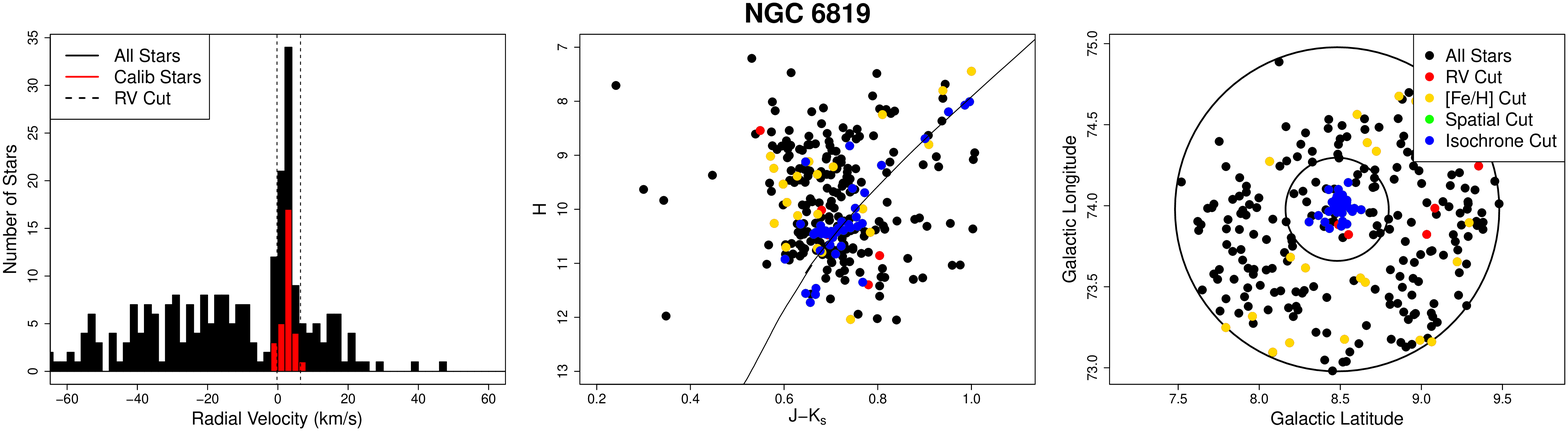}
		\label{fig:n6819_rv_cmd}
    \end{subfigure}%
    ~ 
    \begin{subfigure}%{0.5\textwidth}
        \centering
		\includegraphics[scale=0.34]{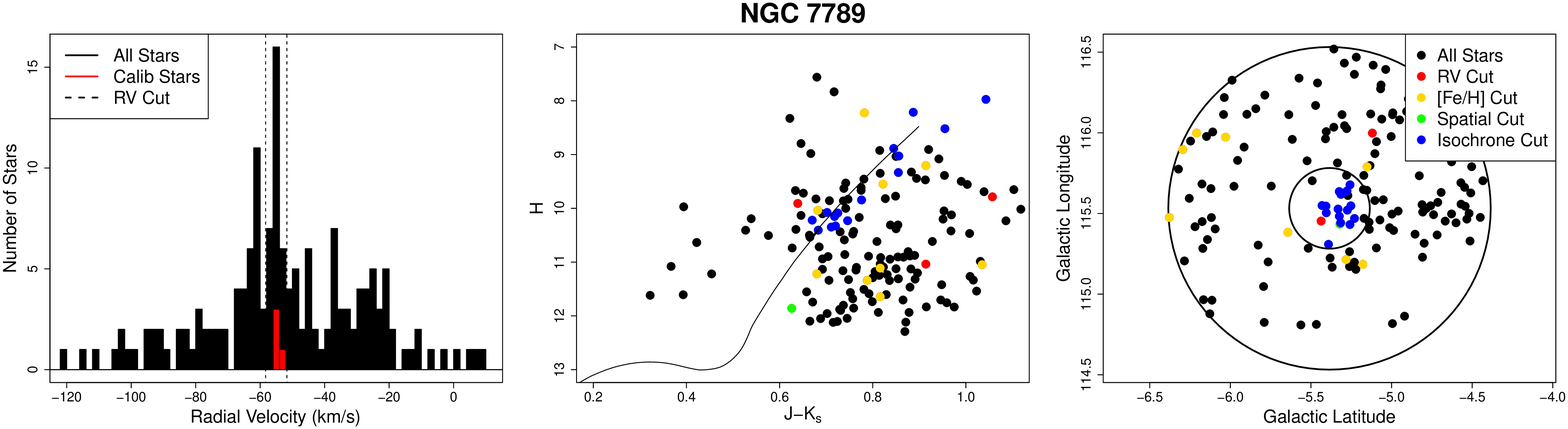}
		\label{fig:n7789_rv_cmd}
    \end{subfigure}
    \caption{Member identification process for other open clusters studied, where colors expressed in the plot are the same as discussed in Figure \ref{fig:n6791_rv_cmd}, with the addition of green representing the additional tidal radius cut when relevant. We overplot all isochrones using the Dartmouth Stellar Evolution Isochrone generator \citep{Dotter et al. 2007} with an [$\alpha$/Fe]$=0.0$. For M67, NGC 188, NGC 6819, and NGC 7789 we used a [Fe/H] value of $-0.01$, $-0.10$, $0.05$, and $0.00$ as well as an age in Gyr of 4.0, 7.0, 2.5, and 1.4 respectively. The dotted lines represent our 1.5$\sigma$ cut on the radial velocity histogram and in $\rm{km\ s}^{-1}$ are centered on 32.5$\pm$1.7, -42.0$\pm$2.0, 3.1$\pm$2.2, and -55.0$\pm$2.2 for the respective clusters.}
\end{figure*}

\begin{figure*}
\label{app:abund}
\centering
\includegraphics[scale=0.65]{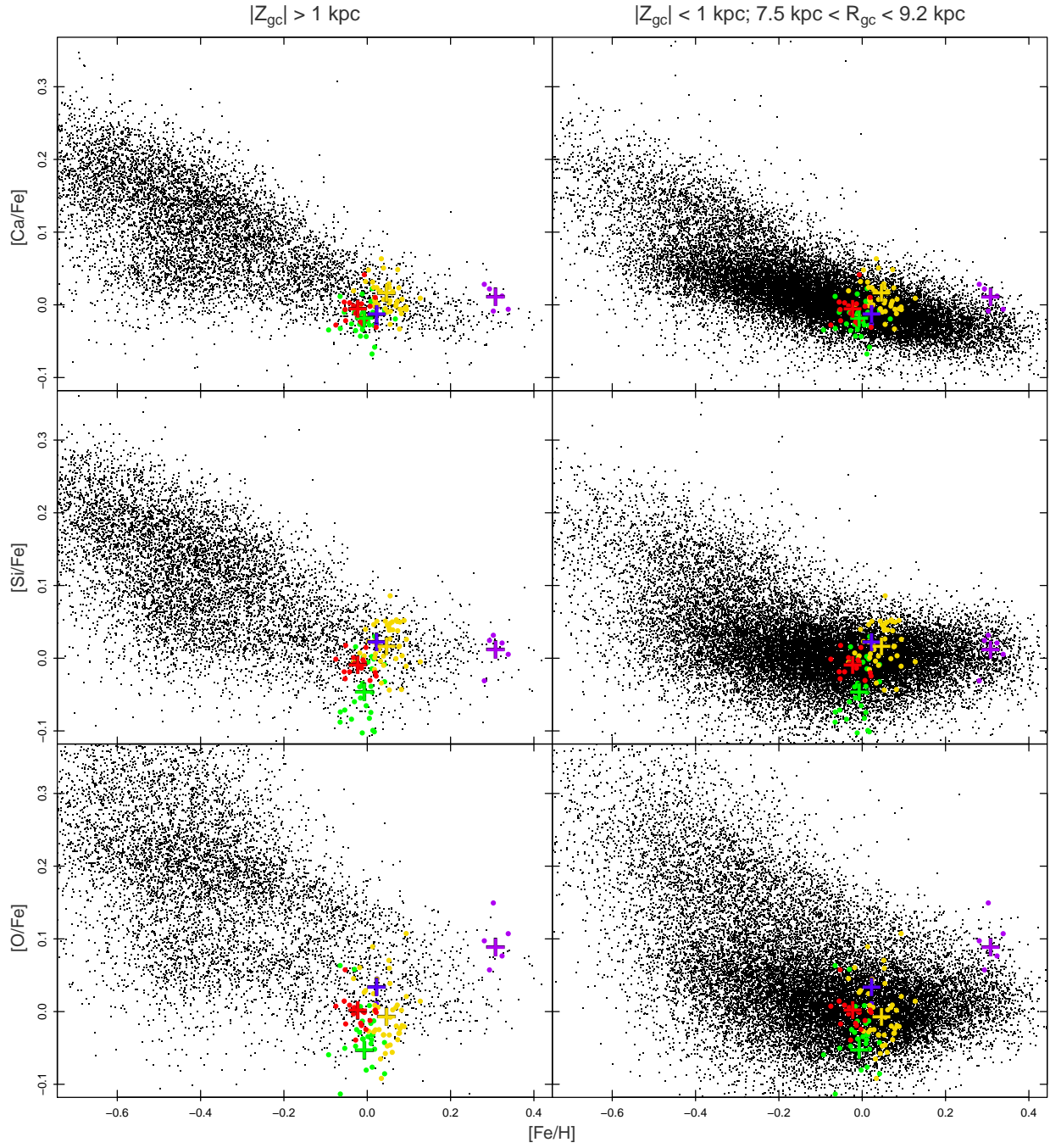}
\caption{Additional abundance plots of relevant elements. The colors for each cluster member follow those adopted in Figure \ref{fig:alpha_seq}.}
\label{fig:multi_plots}
\end{figure*}

We can further confirm the membership of these four younger clusters in the low-sequence by examining their distribution relative to MW field stars for individual elemental abundances (shown in Figure \ref{fig:mg_ca}). Magnesium, carbon, and nitrogen are three of the most reliable and well-studied elements within APOGEE, and Mg is particularly effective in cleanly separating the field stars
into the high- and low-$\alpha$ sequences (Hayes et al. 2017). Figure \ref{fig:mg_ca}, which shows the [Mg/Fe]-[Fe/H] and [(C+N)/Fe]-[Fe/H] distributions of our open cluster candidate stars against those of the field star populations for the high- and low-$Z_{GC}$ sub-samples, respectively, further verifies that the youngest open clusters are consistent with the low-$\alpha$ thin-disk sequence across multiple elemental tracers. These are the chemical distributions where the separation of the sequences is most obvious. However, for completeness we provide in Figure \ref{app:abund} additional $\alpha$-abundance plots for [O/Fe], [Si/Fe], and [Ca/Fe] versus [Fe/H] for all five clusters against MW field stars, and similarly divided into the high- and low-$Z_{GC}$ sub-samples, even though these chemical planes are nowhere near as discriminating between the low- and high-alpha sequences.

For all of the chemical abundance distributions shown in this paper, among the five clusters studied here, NGC 6791 is conspicuously an outlier. The separation seen in the [$\alpha$/Fe]-[Fe/H] plane for stars in NGC 6791 compared to stars in the other four clusters is larger than the abundance \textit{spread} in any particular cluster, and by the \textit{entire spread} of all four clusters in the low alpha sequence. This separation is further confirmed in several of the individual $\alpha$-elements, as demonstrated in Figures \ref{fig:mg_ca} and \ref{fig:multi_plots}. As shown in Table 1, among the five clusters studied here, NGC 6791 is the oldest ($8 \pm 1$ Gyr)\footnote{The age of NGC 6791 varies depending on the model isochrones used. \citep{Grundahl et al. 2008} reported an age 7.7, 8.2, or 9.0 Gyr depending on model isochrones used.}, the most metal-rich ($[Fe/H] = 0.38$), and lies farthest from the Galactic plane at $Z_{GC} = 1.1$ kpc \citep{Carretta et al. 2007, Cunha et al. 2015, Dalessandro et al. 2015}. The fact that NGC 6791 lies nearly four old thin-disk scale heights ($h_{z,D} = 350$ pc) from the Galactic mid-plane already suggests that NGC 6791 is not likely part of the thin disk \citep{Buser 2000}. Our multi-element chemical analysis using APOGEE further supports the peculiarity of this cluster.

\section{Discussion}

In the following sections we describe different scenarios for the origin of NGC 6791, and discuss what implications these have for the evolution of the stellar populations of the MW. In \S \ref{sec:4.1} we present several origins we believe we can rule out based on currently available data. In \S \ref{sec:4.2} we discuss the two most likely remaining possibilities for the origin of NGC 6791 — i.e. that it originates from either the bulge or the thick disk of the MW. In either case, and taking advantage of the universal nature of the high-alpha sequence across the Galaxy, in \S \ref{sec:4.3} we use the five clusters in our sample to time-stamp what appears to be a major transition point (from the high- to low-$\alpha$ sequence) in the chemical evolution of the MW.

\subsection{Unlikely Origins for NGC 6791}
\label{sec:4.1}

Initial examination of the chemistry of NGC 6791 in Figure 2 may lead one to conclude that this cluster is equally misfit from either the low-$\alpha$ or high-$\alpha$ sequence, or that is could lie on an extrapolation on the trends of either sequence. Indeed, Figure 2 conveys the impression that NGC 6791 may actually fit better to an extrapolated low-$\alpha$ sequence than to an extrapolated high-$\alpha$ sequence. Of the various chemical planes shown in Figures 2, 4, and 6, the C+N and Mg distributions have the clearest separation between the low- and high-$\alpha$ sequences and show a prominent feature in the solar-circle cut (right hand panels) that extends from the low-$\alpha$ sequence to the high-metallicity location of NGC 6791. Such a connection to the thick disk sequence (left hand panels) is not as obvious in those chemical planes. This might lead one to believe that this cluster is a super-solar extension of the thin-disk population. However, the shear number of stars in the solar circle cut (51948) versus the thick-disk cut (8600) gives a misleading impression, as the wings of the low-$\alpha$ distribution are much easier to see when you display over six times more stars. In addition, if we compare thin-disk stars of similar age and $\alpha$ abundance to NGC 6791, we find that their median distance ($R_{GC}$) is 2-3 kpc closer to the Galactic center than the present location of NGC 6791 (see Figure 1 of \citealt{Martig et al. 2016}).

Thus, if one were to argue a thin disk origin for NGC 6791, its current Galactic location would need to be explained, given that the chemical counterparts in the thin disk are both closer to the mid-plane and closer to the Galactic center. While, the present $Z_{GC}$ of NGC 6791 could have derived from vertical kinematical heating over a long timescale while orbiting in the thin-disk\footnote{Multiple studies have found that older clusters exhibit a velocity dispersion similar to thin-disk stars of older ages \citep{Carney et al. 1989, Friel 1995}.}, it must be accompanied by an outward radial migration several times larger. \citet{Wu et al. 2009} analyzed the orbits and kinematics of 488 open clusters in the MW and found that the observed radial metallicity gradient for open clusters currently located at $R_{GC} \leq 13.5$ kpc is equivalent to the one derived from the apogalacticon distances of the cluster orbits. This implies that orbits of open clusters are not significantly affected by radial migration, because one would expect the metallicity gradient to flatten out with time \citep{Jacobson et al. 2011}. Thus any radial migration on the order of several kpc, needed to explain the current location of NGC 6791, seems unlikely.

Yet another explanation for NGC 6791's peculiar properties is that it may be the remnant of a tidally disrupted infalling satellite. This would be consistent with the cluster's presently high mass ($\sim 5 \times 10^{3}\ M_{\odot}$) among open clusters. However, while we cannot completely rule this possibility out, we believe it is unlikely given that all dwarf galaxies analyzed to date have an old ($\sim 13$ Gyr) stellar population clearly not present in this cluster \citep{Weisz et al. 2014}. Furthermore, its high metallicity compared with the much lower mean metallicities of Local Group dwarf galaxies \citep{Tolstoy et al. 2009}, and simulations of the chemical evolution of satellites in MW-like halos \citep{Lee et al. 2015} make this possibility even more unlikely. Moreover, the observed stellar velocity dispersion of our member stars (2.5 km s$^{-1}$) is much smaller than typically seen in dark-matter-dominated dwarf satellites \citep{Matteucci 2014}.

If not a dwarf galaxy remnant, perhaps NGC 6791 at least started life as a different kind of cluster. While open clusters are normally associated with the younger, thin disk, NGC 6791 would not violate this paradigm if it were not, in fact, an open cluster to begin with. The idea that NGC 6791 may have started life as a globular cluster has been discussed previously, based on the fact that, unlike other open clusters but similiar to typical globular clusters, NGC 6791 may contain multiple stellar populations, as suggested by an apparent bimodality in the Na abundances of the stars detected by \citet{Geisler et al. 2012}. On the other hand, several follow-up studies to that study have not confirmed the existence of a double Na abundance in NGC 6791, which weakens this chemical argument for its origin as a globular cluster (\citealt{Bragaglia et al. 2014, Boesgaard et al. 2015, Cunha et al. 2015}).

%based not only on an orbital analysis of the system \citep{Jilkova et al. 2012},

\subsection{More Viable Possibilities for NGC 6791’s Origin}
\label{sec:4.2}

Two more compelling explanations for NGC 6791 are that it was born in an already vertically extended disk structure, or that it was expelled as part of a major perturbation from the bulge, likely caused by the bar of the MW. We note that, in the latter case, any major event strong enough to effect NGC 6791 might have also affected much of the rest of the existing disk at that time --- something that may explain the chemical uniformity of the high-$\alpha$ population across the Galaxy. We consider each of these possibilities in the following sections as equally plausible and likely scenarios.

\subsubsection{NGC 6791's Origin as a Bulge Cluster}

Recently, evidence from APOGEE has pointed towards chemical differences in bulge stars identified in Baade's Window (BW) compared with 2904 thin- and thick-disk stars having $R_{GC} \leq 7.7$ kpc from the same data release (DR13) and calibration pipeline (ASPCAP) \citep{Schultheis et al. 2017}. The position of NGC 6791 in the [$\alpha$/Fe]-[Fe/H] plane lies right on the fiducial line for these Baade's Window stars. In fact, we find the agreement of NGC 6791's chemistry to the Baade's Window high-$\alpha$ sequence to be even better than to the nearby thick disk (Figure \ref{fig:alpha_seq}). Thus despite the difficulty in modeling exactly how NGC 6791 arrived at its present location, the detailed chemistry of the cluster strongly suggests a connection with the MW bulge.

Perhaps, then, NGC 6791 formed originally in the inner Galaxy or bulge and then was subsequently moved outwards to its present Galactocentric radius via an interaction with the bar. \citet{Bedin et al. 2006} used the proper motion of NGC 6791 as derived from HST measurements of its stars to determine that this was a plausible scenario for this cluster. More recently, \citet{Jilkova et al. 2012} studied this possibility in detail by modeling the orbit of open clusters in response to a strong tidal perturbation from a galactic bar like that seen in the MW. They find that, under certain physical conditions and assumptions, a strong bar could indeed perturb a cluster to an orbit that matches the apo-Galacticon and peri-Galacticon (8.5 and 5 kpc respectively) derived for NGC 6791 \citep{Jilkova et al. 2012}. Crucially, however, these authors note that the possibility of this actually happening in their simulations was only 0.4\%. Moreover, the parameters of the model capable of displacing NGC 6791 to its present location require us to redistribute the entire mass budget of the bulge in the model to the bar (effectively doubling its mass). Even with this unphysical and unlikely mass redistribution, this model still struggles to reproduce NGC 6791's height above the Galactic mid-plane. Therefore, despite the apparently promising chemical connection between NGC 6791 and the Galactic bulge, it seems challenging to come up with a dynamical scenario that accounts for the cluster's current Galactic location far from the MW center.

\subsubsection{NGC 6791's Origin as a Thick Disk Cluster}

We believe that the large vertical distance from the disk is suggestive of NGC 6791's origin and membership in the thick disk, due to the statistical unlikelihood of being jettisoned from the bulge or displaced by a massive disk heating event. Additionally, this cluster has an age consistent with the distribution of ages seen in the thick-disk ($\sim 8-12$ Gyr), further justifying our use of NGC 6791 below in age-dating the metal-rich end of the high-$\alpha$ sequence (\citealt{Haywood et al. 2013, Bensby et al. 2014}). An important clue to NGC 6791’s origin must be its age. Despite a variety of different means to ascertain its age, all studies agree that NGC 6791 is rather ancient, with derived ages falling in the range from 7-9 Gyr. If one assumes an age at the upper end of this distribution, then NGC 6791 would be a contemporary of at least some MW globular clusters \citep{Grundahl et al. 2008}. Nevertheless, NGC 6791 is more metal rich than GC's at similar ages in the thick disk and halo of the MW that also lie in the annulus $Z \geq 1$ kpc and $8 \leq R_{GC} \le 13$ kpc, which shows that NGC 6791 does not fit in the classical definition of MW GCs, and, at best would be an extreme outlier in the MW GC population. If instead we assume an age for NGC 6791 on the lower end of the 7-9 Gyr range, we can more easily explain it as an open cluster, but one with a unique chemistry (i.e., star formation history) compared to the majority of open clusters found in the thin disk of the MW.

Note that, as mentioned above and shown by \citet{Nidever et al. 2014} and \citet{Hayden et al. 2015}, the shape and distribution of stars along the high-$\alpha$ sequence is independent of position within the disk --- only the relative density of the high-$\alpha$ population with respect to the low-$\alpha$ population varies with Galactocentric radius and distance from the Galactic plane. Thus, the position occupied by NGC 6791 in the high-$\alpha$, high-$[Fe/H]$ region should not depend on the Galactocentric position of this cluster. Given that it is expected that chemical evolution proceeds from low to high metallicity, NGC 6791's unique position suggests that it may have formed at an epoch when that population reached its highest metallicity. Thus, the $\sim 8$ Gyr age for NGC 6791 may put a unique time stamp on the end of the formation of the high-$\alpha$ stellar population of the MW.

%\textbf{We have already discussed the notion that chemically, NGC 6791 appears to be more consistent with bulge stars from Baade's window (\citet{Schultheis et al. 2017} - Figures 13-15). This result, however, has not been seen as clearly in other works. \citet{Bensby et al. 2017} found that when examining micro-lensed giant stars in the bulge, the individual elemental abundances of Mg, O, Ca, and Si, are consistent with both the thick-disk and the high-metallicity end of the thin-disk populations we observe in the Solar Neighborhood. This comes with the caveat that if there is indeed a thin-disk component in the bulge it is far less significant in overall number compared to what has been observed for the thin-disk in the solar neighborhood (\citealt{Adibekyan et al. 2011, Bensby et al. 2014}).}

\subsection{Time Stamps on Galactic Chemical Evolution}
\label{sec:4.3}

While bulge and thick disk stars in the solar neighborhood likely have some differences in their detailed chemistry (e.g., \citealt{Schultheis et al. 2017}), the two populations nevertheless experienced similar, though not identical, chemical evolution \citep{Melendez et al. 2008}. \citet{Bensby et al. 2017} point out that the Galactic bulge has both age and abundance properties that appear to be tightly connected to the other primary Galactic stellar populations. Here we argue that if the bulge and thick disk populations have similar star formation timescales, star formation rates, and initial mass functions, then regardless of which stellar population NGC 6791 came from, the cluster appears to hold a unique role among the presently known and characterized MW star clusters in placing a key time-stamp on a potentially significant phase in the evolution of our Galaxy's stellar populations.

In a study of the evolution of disk-like galaxies, \citet{Leroy et al. 2008} found that regions of the ISM dominated by neutral atomic hydrogen have a star formation efficiency (SFE: ${\rm SFR} (M_{\odot}/{\rm yr}) / M_{*}$) that decreases with radius. If this scenario holds, then early on the entire MW disk must have been dominated by molecular gas, producing a high and roughly constant SFE, the latter explaining the MW-wide uniformity in the observed high-$\alpha$ sequence of stars. Gradually, the ISM began to transition to a neutral atomic state with the infall of pristine gas, reducing its SFE substantially, while the SFE in the inner Galaxy remained high. This net evolutionary scenario would produce a single high-$\alpha$, high-SFE sequence in the inner Galaxy, but a dual-SFE, dual-$\alpha$ sequence in the outer Galaxy, as is observed (e.g. \citet{Hayden et al. 2015}).

In this context, the age ($8 \pm 1$ Gyr) and metallicity of NGC 6791 suggest that this star cluster formed at the very end of the high SFE (high-$\alpha$) sequence, before the infall of pristine gas onto the galaxy caused the jump to the low SFE (low-$\alpha$) sequence, which is at least as old as $\sim 7$ Gyr --- i.e., the age of the cluster NGC 188\footnote{The derived age of NGC 118 varies by $\sim 0.5$ Gyr depending on the adopted value (e.g., Bonatto et al. 2005: 7.0 Gyr, Fornal et al. 2007: 7.5 Gyr, and Meibom et al. 2009: 6.2 Gyr).}, which lies squarely on the low-$\alpha$ sequence.
Remarkably, the evolution from the high-$\alpha$ to low-$\alpha$ chemical sequence appears to be bracketed by the ages of these two clusters, which are currently dated with reasonably accurate confidence to be within 1 Gyr of each other.
\textit{In this way, NGC 6791 and NGC 188 may be a critical pair of star clusters that play a key role in age dating, rather precisely, both the timing and the duration of an important transition point in the chemical history of the MW.} Even more remarkably, this chemical transition age --- i.e., between about 7 and 8 Gyr ago --- is consistent with the one estimated independently by infall models of Galactic chemical evolution ---  where the thin-disk (low-$\alpha$ sequence) was formed with a low SFE and a long timescale ($\sim 8$ Gyr) infall chosen to match the timescales used in \citet{Chiappini et al. 1997}. The model is initially described by \citet{Nidever et al. 2014}, and more fully described in \citet{Andrews et al. 2016}.
 
On the other hand, \citet{Haywood et al. 2013} --- who analyzed a sample of 1111 solar neighborhood stars from \citet{Adibekyan et al. 2011} that have high-quality abundance determinations and $S/N \geq 200$ for most (55\%) of sample --- found a distinct overlap between the ages of stars in the metal-rich, high-$\alpha$ sequence and the metal-poor, low-$\alpha$ sequence in the solar neighborhood. This complicates our view of a sharp transition of the SFE at $\sim 8$ Gyr ago, and implies a more gradual transition between the two modes of disk formation of the MW. While the overlap might be explained by uncertainties in the derived ages for the field-star sample, it still complicates the SFE-transition scenario proposed by \citet{Nidever et al. 2014}. 
%This suggests that the gas which formed the youngest high-$\alpha$ stars set the initial conditions from which the oldest low-$\alpha$ stars started to form.

Further, \citet{Nidever et al. 2014} propose a scenario whereby the bi-modality and the low-$\alpha$ sequence we observe in the solar neighborhood could result from a superposition of stellar populations born at different radii. This model is attractive because it does not require any sharp SFE transitions or fine-tuned infall scenarios to explain the observed sequences. Additionally, the superposition model can explain the shift in the median [Fe/H] of the low-$\alpha$ locus with Galactocentric radius. In this context, the observed age gradient of the four open clusters in our sample would either represent open clusters that formed with four different star formation enrichment histories, or open clusters that formed with a single SFH and subsequently radially migrated. Given the age of the open clusters in our study, we cannot rule out either possibility, even if our observed age gradient is in the opposite sense to what others have found in the solar neighborhood \citep{Frinchaboy et al. 2013}.

Finally, while we have demonstrated how open clusters can be useful for age-dating critical chemical-evolution phases in the disk, this discussion has been limited to the open clusters observed in a homogeneous fashion and in the greatest detail by APOGEE --- a sample selected to ensure the greatest reliability in the inter-comparison of relative cluster chemistries.  This sample will be improved and expanded in the future with APOGEE-2 (including clusters in both the Northern and Southern Hemispheres), and independent analyses of these and other clusters will be provided by other surveys, such as GALAH, LAMOST, and the Gaia ESO survey. With greater coverage of clusters around the Galactic disk, and of the age-chemistry distribution at different points throughout the disk, a confidently time-stamped chemical evolution scenario can be achieved.
%SRM5: We might want to mention asteroseismic ages here as the great new thing, but I think -- we should check -- that those ages are ultimately calibrated upon the cluster timescale?

\section{Conclusion}
We have used elemental-abundance information on Galactic red giant stars in five open clusters from the Apache Point Observatory Galactic Evolution Experiment (APOGEE) DR13 dataset to age-date the chemical evolution of the high- and low-$\alpha$ element sequences of the MW disk system. The following conclusions are reached:

\noindent
(1) NGC 6791's peculiar age, Galactocentric radius, and height above the Galactic plane already mark it as an unusual system, an extreme object with characteristics well outside the norm of typical thin disk open clusters. The APOGEE-derived abundances for stars in NGC 6791 put them in the high metallicity, high-$\alpha$ region of the [$\alpha$/Fe]-[Fe/H] plane, regions also occupied by stars from the Galactic thick disk as well as stars from Baade's Window. This suggests that, at least chemically, NGC 6791 may either have a close association with the thick disk or that it has been displaced from the Galactic bulge. While the NGC 6791 chemical match is closer to Baade's Window stars than for thick disk stars, it is difficult to understand how a cluster formerly in the Galactic bulge could have been ejected to the solar neighborhood. We rule out that this cluster could be a high-$\alpha$ metallicity extension of the low-$\alpha$, thin disk sequence, because of age and metallicity mismatches for field stars at the same galactocentric radius.

\noindent
(2) NGC 6791 lies at the high-metallicity end ([Fe/H] $\sim 0.3$) of the high-$\alpha$ sequence, it allows us to use the derived age of the cluster to place an upper limit to the age of the high metallicity endpoint of the high-$\alpha$ sequence. The age of NGC 6791 also well-matches the transition time adopted in models attempting to explain the evolution from the high- to low-$\alpha$ sequences of the Galactic disk using simple models gas infall onto galaxies. 

\noindent
(3) There is a gap of $\sim 1$ Gyr between NGC 6791 and the next youngest cluster in our sample, NGC 188. The age and chemical location of NGC 188 also allows us to set a lower limit of $7 \pm 0.5$ Gyr on the oldest stars in the low-$\alpha$ sequence of the MW. In this context, NGC 6791 and NGC 188 are potentially a pair of star clusters that bracket both the timing and the duration of an important transition point in the chemical history of the MW.

%\section{Acknowledgments}
\acknowledgements
Funding for the Sloan Digital Sky Survey IV has been provided by
the Alfred P. Sloan Foundation, the U.S. Department of Energy Office of
Science, and the Participating Institutions. SDSS-IV acknowledges
support and resources from the Center for High-Performance Computing at
the University of Utah. The SDSS web site is www.sdss.org.

SDSS-IV is managed by the Astrophysical Research Consortium for the 
Participating Institutions of the SDSS Collaboration including the 
Brazilian Participation Group, the Carnegie Institution for Science, 
Carnegie Mellon University, the Chilean Participation Group, the French Participation Group, Harvard-Smithsonian Center for Astrophysics, 
Instituto de Astrof\'isica de Canarias, The Johns Hopkins University, 
Kavli Institute for the Physics and Mathematics of the Universe (IPMU) / 
University of Tokyo, Lawrence Berkeley National Laboratory, 
Leibniz Institut f\"ur Astrophysik Potsdam (AIP),  
Max-Planck-Institut f\"ur Astronomie (MPIA Heidelberg), 
Max-Planck-Institut f\"ur Astrophysik (MPA Garching), 
Max-Planck-Institut f\"ur Extraterrestrische Physik (MPE), 
National Astronomical Observatory of China, New Mexico State University, 
New York University, University of Notre Dame, 
Observat\'ario Nacional / MCTI, The Ohio State University, 
Pennsylvania State University, Shanghai Astronomical Observatory, 
United Kingdom Participation Group,
Universidad Nacional Aut\'onoma de M\'exico, University of Arizona, 
University of Colorado Boulder, University of Oxford, University of Portsmouth, 
University of Utah, University of Virginia, University of Washington, University of Wisconsin, 
Vanderbilt University, and Yale University.

R.C. acknowledges support provided by the Spanish Ministry of Economy and Competiviness under grants AYA2013-42781-P, AYA2014-56795-P, and AYA2014-56359-P. T.C.B. acknowledges partial support for this work from the National Science Foundation under Grant No. PHY 14-30152 (JINA Center for the Evolution of the Elements). J.G.F-T acknowledges support provided by Centre National d\'Etudes Spatiales (CNES) and the R\'egion de Franche-Comt\'e and by the French Programme National de Cosmologie et Galaxies (PNCG). D.G., B.T. and S.V. gratefully acknowledge support from the Chilean BASAL Centro de Excelencia en Astrof\'isica y Tecnolog\'ias Afines (CATA) grant PFB-06/2007. PMF  acknowledges support for this work from the National Science Foundation under Grant No. AST-1311835. S.R.M acknowledges support from National Science Foundation. We thank the referee for helpful comments that improved this paper.

\begin{deluxetable*}{cccccccccccccc}
\label{tab:table}
\center
\tabletypesize{\footnotesize}
\tablecolumns{13}
\tablewidth{0pt}
\tablecaption{Individual Cluster Members Properties.}
\tablehead{
\colhead{Cluster} & \colhead{2MASS ID} & \colhead{$J-K_s$} & \colhead{$H$} & \colhead{$T_{\rm eff}$} & \colhead{log g} & \colhead{$V_r$} & \colhead{[$\alpha$/Fe]} & \colhead{[Fe/H]} & \colhead{[Mg/Fe]} & \colhead{[(C+N)/Fe]} & \colhead{[Ca/Fe]} & \colhead{[Si/Fe]} & \colhead{[O/Fe]} 
\\
\colhead{} & \colhead{} & \colhead{} & \colhead{} & \colhead{(K)} & \colhead{} & \colhead{(km/s)} & \colhead{} & \colhead{} & \colhead{} & \colhead{} & \colhead{} & \colhead{} & \colhead{}
}
\startdata
NGC 188  & J00533497+8511145 & 0.67 & 10.79 & 4664 & 2.80 & -42.1       & 0.011     & 0.022  & 0.044  & 0.127  & -0.026 & 0.023  & 0.011  \\
NGC 188  & J01025280+8517563 & 0.80 & 9.84  & 4470 & 2.41 & -42.1       & 0.033     & 0.023  & 0.053  & 0.127  & -0.003 & 0.020  & 0.045  \\
M67      & J08511704+1150464 & 0.68 & 8.71  & 4730 & 2.74 & 33.5        & -0.021    & -0.026 & 0.010  & 0.050  & -0.007 & -0.005 & -0.025 \\
M67      & J08511564+1150561 & 0.47 & 11.09 & 5243 & 3.70 & 34.5        & -0.023    & -0.031 & -0.048 & -0.017 & -0.011 & -0.060 & 0.058  \\
M67      & J08511877+1151186 & 0.48 & 11.09 & 5200 & 3.72 & 34.3        & -0.012    & 0.009  & -0.038 & -0.073 & 0.005  & -0.075 & -0.016 \\
M67      & J08512156+1146061 & 0.66 & 9.09  & 4758 & 2.87 & 34.7        & -0.011    & 0.041  & 0.014  & 0.045  & -0.007 & -0.009 & -0.043 \\
M67      & J08512898+1150330 & 0.61 & 8.07  & 4713 & 2.46 & 33.4        & -0.032    & 0.011  & -0.004 & 0.062  & -0.068 & -0.014 & -0.032 \\
M67      & J08510839+1147121 & 0.58 & 10.19 & 4930 & 3.34 & 33.6        & -0.027    & 0.042  & -0.028 & 0.003  & -0.010 & -0.032 & -0.086 \\
M67      & J08512879+1151599 & 0.41 & 11.10 & 5639 & 3.92 & 33.4        & -0.056    & -0.093 & -0.071 & 0.007  & -0.035 & -0.124 & -0.059 \\
M67      & J08512935+1145275 & 0.53 & 10.86 & 5049 & 3.63 & 33.1        & -0.050    & -0.012 & -0.036 & -0.018 & 0.015  & -0.103 & -0.209 \\
M67      & J08511269+1152423 & 0.67 & 8.12  & 4761 & 2.48 & 34.4        & -0.013    & 0.002  & 0.001  & 0.052  & -0.023 & -0.005 & -0.034 \\
M67      & J08510106+1150108 & 0.43 & 11.02 & 5438 & 3.94 & 32.9        & -0.052    & -0.055 & -0.077 & -0.020 & -0.010 & -0.124 & -0.181 \\
M67      & J08512618+1153520 & 0.66 & 8.11  & 4777 & 2.51 & 34.1        & -0.022    & -0.010 & -0.009 & 0.058  & -0.024 & 0.001  & -0.038 \\
M67      & J08513938+1151456 & 0.59 & 9.89  & 4879 & 3.22 & 34.4        & -0.021    & -0.012 & -0.021 & 0.031  & -0.034 & -0.038 & -0.013 \\
M67      & J08514234+1150076 & 0.64 & 9.34  & 4769 & 2.96 & 34.3        & -0.011    & 0.013  & 0.043  & 0.040  & -0.010 & -0.099 & -0.031 \\
M67      & J08514235+1151230 & 0.64 & 8.85  & 4727 & 2.81 & 33.5        & -0.019    & -0.011 & 0.006  & 0.059  & -0.030 & -0.035 & -0.038 \\
M67      & J08514507+1147459 & 0.64 & 9.18  & 4773 & 2.91 & 32.9        & -0.041    & 0.012  & 0.014  & 0.018  & 0.022  & -0.047 & -0.076 \\
M67      & J08513577+1153347 & 0.58 & 10.02 & 4887 & 3.23 & 34.1        & -0.043    & -0.003 & -0.005 & 0.000  & -0.044 & -0.065 & -0.081 \\
M67      & J08514401+1146245 & 0.41 & 11.11 & 5476 & 3.93 & 33.0        & -0.084    & -0.016 & -0.086 & -0.027 & -0.043 & -0.148 & -0.249 \\
M67      & J08514474+1146460 & 0.54 & 10.92 & 5059 & 3.64 & 33.5        & -0.016    & -0.054 & -0.029 & 0.011  & -0.007 & -0.071 & -0.050 \\
M67      & J08505816+1152223 & 0.57 & 10.71 & 4968 & 3.52 & 34.1        & -0.026    & 0.009  & -0.039 & -0.045 & -0.001 & -0.039 & -0.044 \\
M67      & J08504994+1149127 & 0.48 & 10.96 & 5158 & 3.70 & 33.7        & -0.044    & -0.065 & -0.055 & -0.053 & -0.033 & -0.074 & -0.113 \\
M67      & J08510951+1141449 & 0.45 & 11.10 & 5459 & 3.90 & 32.3        & -0.049    & 0.017  & -0.069 & -0.022 & 0.005  & -0.101 & -0.213 \\
M67      & J08510018+1154321 & 0.46 & 10.92 & 5308 & 3.82 & 34.1        & -0.030    & 0.005  & -0.060 & -0.098 & -0.028 & -0.083 & 0.008  \\
M67      & J08515611+1150147 & 0.56 & 10.73 & 4929 & 3.55 & 34.7        & -0.066    & 0.067  & -0.027 & -0.058 & -0.019 & -0.157 & -0.161 \\
M67      & J08511897+1158110 & 0.57 & 10.10 & 4909 & 3.27 & 34.0        & -0.027    & -0.018 & -0.007 & 0.021  & -0.027 & -0.039 & -0.029 \\
M67      & J08514388+1156425 & 0.62 & 8.11  & 4757 & 2.47 & 32.9        & -0.026    & 0.011  & 0.005  & 0.037  & -0.015 & 0.002  & -0.039 \\
M67      & J08513540+1157564 & 0.42 & 11.14 & 5469 & 3.80 & 33.7        & -0.014    & -0.065 & -0.063 & -0.008 & 0.012  & -0.088 & 0.063  \\
M67      & J08514883+1156511 & 0.55 & 10.78 & 4971 & 3.48 & 34.3        & -0.039    & -0.038 & 0.005  & -0.004 & -0.026 & -0.084 & -0.161 \\
M67      & J08515952+1155049 & 0.64 & 8.08  & 4723 & 2.40 & 34.3        & -0.030    & -0.028 & -0.005 & 0.043  & -0.035 & 0.016  & -0.048 \\
M67      & J08503613+1143180 & 0.58 & 10.64 & 5008 & 3.56 & 34.0        & -0.011    & -0.016 & -0.017 & 0.029  & -0.015 & -0.055 & 0.007  \\
M67      & J08521134+1145380 & 0.46 & 11.08 & 5336 & 3.78 & 33.1        & -0.070    & -0.011 & -0.078 & -0.078 & -0.011 & -0.126 & -0.308 \\
M67      & J08521856+1144263 & 0.65 & 8.09  & 4727 & 2.47 & 33.6        & -0.025    & 0.018  & -0.003 & 0.059  & -0.058 & 0.030  & -0.052 \\
M67      & J08504964+1135089 & 0.69 & 8.85  & 4750 & 2.80 & 34.8        & -0.001    & 0.016  & 0.010  & 0.055  & -0.028 & -0.001 & -0.013 \\
NGC 6791 & J19205338+3748282 & 0.96 & 9.96  & 4027 & 1.63 & -48.5       & 0.096     & 0.324  & 0.117  & 0.195  & 0.012  & 0.021  & 0.077  \\
NGC 6791 & J19210426+3747187 & 0.91 & 10.31 & 4115 & 1.71 & -46.8       & 0.080     & 0.293  & 0.096  & 0.179  & 0.022  & 0.024  & 0.057  \\
NGC 6791 & J19205530+3743152 & 0.86 & 11.27 & 4261 & 2.22 & -47.9       & 0.098     & 0.281  & 0.114  & 0.212  & 0.028  & -0.031 & 0.097  \\
NGC 6791 & J19210112+3742134 & 0.88 & 11.13 & 4227 & 2.23 & -47.7       & 0.100     & 0.338  & 0.145  & 0.210  & -0.006 & 0.005  & 0.107  \\
NGC 6791 & J19211007+3750008 & 0.81 & 12.11 & 4468 & 2.63 & -49.0       & 0.087     & 0.303  & 0.146  & 0.209  & -0.009 & 0.032  & 0.149  \\
NGC 6819 & J19411705+4010517 & 1.00 & 8.01  & 4062 & 1.49 & 1.4         & 0.015     & 0.017  & -0.007 & 0.078  & 0.010  & -0.016 & 0.001  \\
NGC 6819 & J19411893+4011408 & 0.72 & 10.68 & 4710 & 2.53 & 1.0         & -0.009    & 0.032  & -0.001 & 0.069  & 0.018  & -0.004 & -0.065 \\
NGC 6819 & J19411476+4011008 & 0.64 & 10.27 & 4858 & 2.57 & 1.1         & 0.002     & 0.078  & 0.003  & 0.045  & -0.006 & 0.050  & -0.023 \\
NGC 6819 & J19412136+4011002 & 0.74 & 10.28 & 4655 & 2.41 & 1.9         & 0.000     & 0.034  & -0.027 & 0.082  & 0.063  & -0.034 & -0.092 \\
NGC 6819 & J19411564+4010105 & 0.70 & 10.52 & 4778 & 2.48 & 1.6         & 0.010     & 0.076  & -0.028 & 0.041  & 0.023  & 0.016  & -0.030 \\
NGC 6819 & J19412176+4012111 & 0.74 & 10.34 & 4628 & 2.44 & 0.8         & -0.006    & 0.069  & 0.010  & 0.053  & -0.002 & -0.001 & 0.006  \\
NGC 6819 & J19411102+4011116 & 0.65 & 9.12  & 4911 & 2.42 & 3.0         & 0.013     & 0.055  & -0.027 & 0.126  & -0.027 & 0.086  & -0.046 \\
NGC 6819 & J19411115+4011422 & 0.74 & 8.83  & 4652 & 2.27 & 4.6         & 0.015     & 0.088  & 0.004  & 0.149  & -0.005 & 0.052  & -0.021 \\
NGC 6819 & J19411355+4012205 & 0.70 & 10.41 & 4802 & 2.51 & 2.5         & -0.015    & 0.016  & -0.006 & 0.051  & -0.001 & 0.046  & -0.028 \\
NGC 6819 & J19411279+4012238 & 0.68 & 10.45 & 4777 & 2.43 & 2.8         & 0.000     & 0.060  & -0.009 & 0.046  & 0.016  & 0.050  & -0.056 \\
NGC 6819 & J19412658+4011418 & 0.81 & 9.19  & 4443 & 2.10 & 2.2         & 0.003     & 0.026  & -0.007 & 0.081  & 0.000  & 0.020  & 0.006  \\
NGC 6819 & J19412369+4012355 & 0.90 & 8.69  & 4234 & 1.82 & 3.8         & 0.010     & 0.018  & -0.002 & 0.076  & 0.006  & 0.001  & 0.009  \\
NGC 6819 & J19412707+4012283 & 0.77 & 9.69  & 4587 & 2.49 & 1.0         & 0.013     & 0.020  & -0.004 & 0.105  & 0.006  & 0.006  & -0.025 \\
NGC 6819 & J19410622+4010532 & 0.67 & 11.46 & 4883 & 2.82 & 3.3         & 0.003     & 0.052  & 0.001  & 0.047  & 0.029  & 0.042  & 0.070  \\
NGC 6819 & J19412953+4012210 & 0.72 & 10.35 & 4755 & 2.47 & 3.1         & -0.017    & 0.077  & -0.004 & 0.053  & -0.033 & 0.017  & -0.030 \\
NGC 6819 & J19412915+4013040 & 0.67 & 10.44 & 4800 & 2.46 & 1.3         & 0.010     & 0.061  & -0.001 & 0.055  & 0.018  & 0.039  & 0.002  \\
NGC 6819 & J19412147+4013573 & 0.69 & 10.40 & 4794 & 2.53 & 1.5         & 0.011     & 0.070  & -0.010 & 0.063  & 0.002  & 0.053  & -0.020 \\
NGC 6819 & J19410858+4013299 & 0.68 & 10.39 & 4769 & 2.50 & 2.5         & -0.006    & 0.048  & -0.006 & 0.051  & -0.010 & 0.046  & -0.047 \\
NGC 6819 & J19413031+4009005 & 0.99 & 8.07  & 4079 & 1.55 & 6.2         & -0.015    & 0.039  & -0.031 & 0.069  & 0.007  & -0.010 & -0.032 \\
NGC 6819 & J19413330+4012349 & 0.76 & 10.31 & 4674 & 2.40 & 4.3         & 0.010     & -0.004 & 0.012  & 0.071  & 0.032  & 0.040  & 0.029  \\
NGC 6819 & J19410524+4014042 & 0.69 & 10.46 & 4823 & 2.53 & 3.2         & 0.013     & 0.073  & -0.014 & 0.083  & 0.048  & 0.038  & -0.039 \\
NGC 6819 & J19412942+4014199 & 0.73 & 10.39 & 4716 & 2.48 & 3.5         & 0.014     & 0.039  & 0.017  & 0.089  & 0.051  & 0.003  & -0.032 \\
NGC 6819 & J19410926+4014436 & 0.66 & 10.45 & 4786 & 2.58 & 2.4         & -0.025    & 0.053  & -0.004 & 0.052  & 0.002  & -0.044 & -0.032 \\
NGC 6819 & J19410819+4015085 & 0.68 & 10.77 & 4788 & 2.75 & 0.2         & 0.020     & -0.033 & -0.009 & 0.261  & 0.019  & -0.006 & 0.045  \\
NGC 6819 & J19413027+4015218 & 0.77 & 10.26 & 4783 & 2.46 & 2.2         & 0.002     & 0.043  & -0.003 & 0.066  & -0.005 & 0.051  & -0.058 \\
NGC 6819 & J19405797+4008174 & 0.68 & 10.30 & 4821 & 2.53 & 4.5         & 0.002     & 0.072  & 0.005  & 0.055  & 0.005  & 0.051  & 0.009  \\
NGC 6819 & J19405601+4013395 & 0.65 & 11.56 & 4908 & 2.88 & 3.6         & -0.034    & 0.013  & -0.024 & 0.042  & 0.031  & -0.170 & 0.089  \\
NGC 6819 & J19410991+4015495 & 0.72 & 10.23 & 4781 & 2.43 & 2.4         & 0.010     & 0.003  & -0.012 & 0.068  & 0.048  & 0.023  & -0.027 \\
NGC 6819 & J19412222+4016442 & 0.71 & 10.43 & 4653 & 2.47 & 2.7         & -0.017    & 0.028  & -0.002 & 0.053  & 0.019  & 0.012  & -0.025 \\
NGC 6819 & J19405020+4013109 & 0.66 & 10.46 & 4756 & 2.57 & 4.4         & -0.009    & 0.082  & -0.011 & 0.043  & -0.012 & -0.043 & 0.002  \\
NGC 6819 & J19404965+4014313 & 0.70 & 10.66 & 4724 & 2.53 & 3.2         & -0.001    & 0.053  & 0.006  & 0.048  & 0.024  & 0.039  & 0.060  \\
NGC 6819 & J19405560+4006292 & 0.75 & 9.61  & 4596 & 2.67 & 2.4         & 0.006     & 0.127  & 0.037  & 0.083  & 0.009  & -0.005 & 0.014  \\
NGC 6819 & J19404803+4008085 & 0.75 & 9.98  & 4561 & 2.44 & 2.3         & 0.016     & 0.023  & 0.017  & 0.084  & 0.024  & 0.051  & 0.045  \\
NGC 6819 & J19412730+4004548 & 0.77 & 11.35 & 4747 & 2.42 & 1.2         & 0.024     & 0.054  & 0.042  & 0.099  & 0.006  & 0.027  & 0.039  \\
NGC 6819 & J19413439+4017482 & 0.95 & 8.20  & 4145 & 1.63 & 2.6         & -0.011    & 0.084  & -0.019 & 0.069  & 0.008  & 0.003  & -0.019 \\
NGC 6819 & J19414427+4005527 & 0.75 & 10.14 & 4606 & 2.42 & 3.8         & 0.018     & 0.015  & 0.022  & 0.086  & 0.024  & 0.022  & 0.025  \\
NGC 6819 & J19411367+4003382 & 0.71 & 10.83 & 4694 & 2.42 & 2.9         & 0.022     & -0.007 & 0.005  & 0.122  & -0.015 & 0.003  & 0.026  \\
NGC 6819 & J19415064+4016010 & 0.66 & 11.73 & 4908 & 2.95 & 4.3         & -0.007    & 0.094  & -0.018 & 0.011  & -0.007 & 0.026  & 0.021  \\
NGC 6819 & J19403684+4015172 & 0.67 & 11.58 & 4911 & 2.91 & 2.3         & -0.021    & 0.093  & -0.020 & 0.014  & -0.003 & -0.011 & 0.107  \\
NGC 6819 & J19412386+4021444 & 0.67 & 10.43 & 4805 & 2.55 & 2.1         & -0.004    & 0.055  & -0.011 & 0.063  & -0.009 & 0.048  & -0.008 \\
NGC 6819 & J19415437+4002097 & 0.60 & 10.92 & 4938 & 2.84 & 4.2         & 0.008     & -0.018 & -0.021 & 0.189  & -0.004 & 0.007  & 0.061  \\
NGC 7789 & J23573184+5641221 & 0.89 & 8.21  & 4378 & 2.03 & -55.6       & -0.017    & -0.016 & -0.031 & 0.058  & -0.023 & -0.009 & -0.016 \\
NGC 7789 & J23571400+5640586 & 0.86 & 9.03  & 4494 & 2.24 & -53.7       & -0.002    & -0.010 & -0.013 & 0.048  & -0.026 & -0.002 & 0.000  \\
NGC 7789 & J23573563+5640000 & 0.67 & 10.22 & 4954 & 2.86 & -55.0       & -0.026    & 0.019  & -0.022 & 0.006  & 0.010  & -0.020 & 0.002  \\
NGC 7789 & J23571728+5645333 & 0.68 & 10.41 & 4973 & 2.93 & -55.7       & -0.040    & -0.007 & -0.038 & -0.006 & 0.041  & -0.006 & -0.001 \\
NGC 7789 & J23565751+5645272 & 0.85 & 9.33  & 4560 & 2.37 & -55.0       & -0.019    & -0.014 & -0.022 & 0.037  & 0.001  & -0.014 & -0.019 \\
NGC 7789 & J23570895+5648504 & 0.71 & 10.35 & 4938 & 2.73 & -54.9       & -0.031    & 0.007  & -0.039 & -0.005 & -0.002 & -0.031 & -0.012 \\
NGC 7789 & J23575438+5647439 & 0.73 & 10.09 & 4956 & 2.77 & -53.4       & -0.009    & -0.033 & -0.022 & 0.063  & -0.001 & 0.001  & -0.003 \\
NGC 7789 & J23563930+5645242 & 0.72 & 10.33 & 4907 & 2.72 & -55.4       & -0.030    & -0.042 & -0.041 & 0.014  & 0.003  & -0.019 & 0.009  \\
NGC 7789 & J23580275+5647208 & 0.78 & 9.84  & 4728 & 2.51 & -54.5       & -0.039    & -0.028 & -0.029 & 0.024  & -0.012 & -0.008 & -0.039 \\
NGC 7789 & J23580275+5647208 & 0.78 & 9.84  & 4753 & 2.55 & -54.9       & -0.002    & -0.053 & -0.012 & 0.058  & -0.006 & -0.028 & 0.057  \\
NGC 7789 & J23571847+5650271 & 0.70 & 10.08 & 4855 & 2.59 & -55.9       & -0.020    & -0.052 & -0.024 & 0.020  & -0.022 & 0.018  & -0.017 \\
NGC 7789 & J23580015+5650125 & 0.95 & 8.52  & 4394 & 2.03 & -54.3       & -0.007    & -0.055 & -0.023 & 0.056  & 0.004  & -0.019 & 0.014  \\
NGC 7789 & J23562953+5648399 & 0.75 & 10.23 & 4884 & 2.66 & -55.3       & -0.018    & -0.075 & -0.044 & 0.037  & -0.028 & -0.001 & 0.007  \\
NGC 7789 & J23564304+5650477 & 0.72 & 10.15 & 4904 & 2.69 & -53.6       & -0.030    & -0.003 & -0.038 & 0.015  & -0.020 & 0.015  & -0.024 \\
NGC 7789 & J23581471+5651466 & 1.04 & 7.97  & 4286 & 1.76 & -56.0       & -0.003    & -0.031 & -0.024 & 0.069  & -0.003 & -0.007 & 0.006  \\
NGC 7789 & J23554966+5639180 & 0.85 & 8.89  & 4466 & 2.15 & -57.0       & -0.032    & 0.021  & -0.006 & 0.015  & -0.031 & -0.025 & -0.001
\enddata
\end{deluxetable*}

\FloatBarrier

\begin{thebibliography}{}

\bibitem[Adibekyan et al.(2011)]{Adibekyan et al. 2011} Adibekyan, V.~Z., Santos, N.~C., Sousa, S.~G., \& Israelian, G.\ 2011, \aap, 535, L11
\bibitem[Albareti et al.(2016)]{Albareti et al. 2016} Albareti, F.~D., Allende Prieto, C., et al. 2016, arXiv:1608.02013
%\bibitem[SDSS Collaboration et al.(2016)]{Albareti et al. 2016} SDSS Collaboration, Albareti, F.~D., Allende Prieto, C., et al.\ 2016, arXiv:1608.02013
\bibitem[Andrews et al.(2016)]{Andrews et al. 2016} Andrews, B.~H., Weinberg, D.~H., Sch{\"o}nrich, R., \& Johnson, J.~A.\ 2016, arXiv:1604.08613 
\bibitem[Angus et al.(2015)]{Angus et al. 2015} Angus, R., Aigrain, S., Foreman-Mackey, D., \& McQuillan, A.\ 2015, \mnras, 450, 1787
\bibitem[Asplund et al.(2005)]{Asplund et al. 2005} Asplund, M., Grevesse, N., \& Sauval, A.~J.\ 2005, Cosmic Abundances as Records of Stellar Evolution and Nucleosynthesis, 336, 25 
\bibitem[Barnes et al.(2016)]{Barnes et al. 2016} Barnes, S.~A., Weingrill, J., Fritzewski, D., Strassmeier, K.~G., \& Platais, I.\ 2016, \apj, 823, 16 
\bibitem[Bedin et al.(2006)]{Bedin et al. 2006} Bedin, L.~R., Piotto, G., Carraro, G., King, I.~R., \& Anderson, J.\ 2006, \aap, 460, L27 
\bibitem[Belokurov et al.(2006)]{Belokurov et al. 2006} Belokurov, V., Zucker, D.~B., Evans, N.~W., et al.\ 2006, \apjl, 642, L137
\bibitem[Bensby et al.(2003)]{Bensby et al. 2003} Bensby, T., Feltzing, S., \& Lundstr{\"o}m, I.\ 2003, \aap, 410, 527
\bibitem[Bensby et al.(2007)]{Bensby et al. 2007} Bensby, T., Zenn, A.~R., Oey, M.~S., \& Feltzing, S.\ 2007, \apjl, 663, L13
\bibitem[Bensby et al.(2014)]{Bensby et al. 2014} Bensby, T., Feltzing, S., \& Oey, M.~S.\ 2014, \aap, 562, A71 
\bibitem[Bensby et al.(2017)]{Bensby et al. 2017} Bensby, T., Feltzing, S., Gould, A., et al.\ 2017, arXiv:1702.02971 
%\bibitem[Blanton et al.(2017)]{Blanton et al. 2017} Blanton, M.R., et al. 2017, AJ, submitted
\bibitem[Blanton et al.(2017)]{Blanton et al. 2017} Blanton, M.~R., Bershady, M.~A., Abolfathi, B., et al.\ 2017, arXiv:1703.00052
\bibitem[Boesgaard et al.(2015)]{Boesgaard et al. 2015} Boesgaard, A.~M., Lum, M.~G., \& Deliyannis, C.~P.\ 2015, \apj, 799, 202
\bibitem[Bonatto et al.(2005)]{Bonatto et al. 2005} Bonatto, C., Bica, E., \& Santos, J.~F.~C., Jr.\ 2005, \aap, 433, 917
\bibitem[Bovy et al.(2012a)]{Bovy et al. 2012a} Bovy, J., Rix, H.-W., \& Hogg, D. W. 2012a, ApJ, 751, 131
\bibitem[Bovy et al.(2012b)]{Bovy et al. 2012b} Bovy, J., Rix, H.-W., Hogg, D. W., et al. 2012b, ApJ, 755, 115
\bibitem[Bovy (2016)]{Bovy 2016} Bovy, J.\ 2016, \apj, 817, 49
\bibitem[Bragaglia et al.(2014)]{Bragaglia et al. 2014} Bragaglia, A., Sneden, C., Carretta, E., et al.\ 2014, \apj, 796, 68
\bibitem[Brogaard et al.(2011)]{Brogaard et al. 2011} Brogaard, K., Bruntt, H., Grundahl, F., et al.\ 2011, \aap, 525, A2
\bibitem[Brook et al.(2004)]{Brook et al. 2004} Brook, C.~B., Kawata, D., \& Gibson, B.~K.\ 2004, Satellites and Tidal Streams, 327, 100
\bibitem[Buser (2000)]{Buser 2000} Buser, R.\ 2000, Science, 287, 69
\bibitem[Cardelli et al.(1989)]{Cardelli et al. 1989} Cardelli, J.~A., Clayton, G.~C., \& Mathis, J.~S.\ 1989, \apj, 345, 245
\bibitem[Carney et al.(1989)]{Carney et al. 1989} Carney, B. W., Latham, D. W. \& Laird, J. B., 1989. Astron. J. 97:423
\bibitem[Carraro \& Chiosi(1994)]{Carraro&Chiosi 1994} Carraro, G. \& Chiosi C. 1994, A\&A, 287, 761
\bibitem[Carraro (2014)]{Carraro 2014} Carraro, G.\ 2014, Tenth Pacific Rim Conference on Stellar Astrophysics, 482, 245
\bibitem[Carrera (2012)]{Carrera 2012} Carrera, R.\ 2012, \apj, 758, 110
\bibitem[Carretta et al.(2007)]{Carretta et al. 2007} Carretta, E., Bragaglia, A., \& Gratton, R. G., 2007, A\&A 473, 129-141
\bibitem[Chiappini et al.(1997)]{Chiappini et al. 1997} Chiappini, C., Matteucci, F. \& Gratton, R. G. 1997, ApJ, 477:765–780
\bibitem[Chiappini et al.(2001)]{Chiappini et al. 2001} Chiappini, C., Matteucci, F., \& Romano, D. 2001, ApJ, 554, 1044
\bibitem[Chiappini et al.(2015)]{Chiappini et al. 2015} Chiappini, C., Anders, F., Rodrigues, T.~S., et al.\ 2015, \aap, 576, L12
\bibitem[Clem et al.(2004)]{Clem et al. 2004} Clem, J. L., VandenBerg, D. A., Grundahl, F., \& Bell, R. A. 2004, AJ, 127, 1227
\bibitem[Cunha et al.(2015)]{Cunha et al. 2015} Cunha, K., Smith, V.~V., Johnson, J.~A., et al.\ 2015, \apjl, 798, L41
\bibitem[Cunha et al.(2016)]{Cunha et al. 2016} Cunha, K., Frinchaboy, P. M., Souto, D. et al. 2016, AN, 337, 922
\bibitem[Dalessandro et al.(2015)]{Dalessandro et al. 2015} Dalessandro, E., Miocchi, P., Carraro, G., J{\'{\i}}lkov{\'a}, L., \& Moitinho, A.\ 2015, \mnras, 449, 1811
\bibitem[De Silva et al.(2015)]{De Silva et al. 2015} De Silva, G.~M., Freeman, K.~C., Bland-Hawthorn, J., et al.\ 2015, \mnras, 449, 2604 
\bibitem[Dias et al.(2002)]{Dias et al. 2002} Dias, W.S., Alessi, B.S., Moitinho, A., Lépine, J.R.D. \& Alessi, B.S. 2002, A\&A, 389, 8718
\bibitem[Dotter et al.(2007)]{Dotter et al. 2007} Dotter, A., Chaboyer, B., Jevremovi{\'c}, D., et al.\ 2007, \aj, 134, 376
\bibitem[Eisenstein et al.(2011)]{Eisenstein et al. 2011} Eisenstein, D. J., Weinberg, D. H., Agol, E., et al. 2011, AJ, 142, 72
\bibitem[Epstein et al.(2014)]{Epstein et al. 2014} Epstein, C.~R., Elsworth, Y.~P., Johnson, J.~A., et al.\ 2014, \apjl, 785, L28
\bibitem[Feuillet et al.(2016)]{Feuillet et al. 2016} Feuillet, D.~K., Bovy, J., Holtzman, J., et al.\ 2016, \apj, 817, 40 
\bibitem[Friel (1989)]{Friel 1989} Friel, E.~D.\ 1989, \pasp, 101, 244
\bibitem[Friel (1995)]{Friel 1995} Friel, E.~D.\ 1995, \araa, 33, 381
\bibitem[Friel (2013)]{Friel 2013} Friel, E.~D.\ 2013, Planets, Stars and Stellar Systems.~Volume 5: Galactic Structure and Stellar Populations, 5, 347
\bibitem[Frinchaboy et al.(2013)]{Frinchaboy et al. 2013} Frinchaboy, P.~M., Thompson, B., Jackson, K.~M., et al.\ 2013, \apjl, 777, L1
\bibitem[Fuhrmann (1998)]{Fuhrmann 1998} Fuhrmann, K.\ 1998, \aap, 338, 161 
\bibitem[Garc{\'{\i}}a P{\'e}rez et al.(2016)]{Garcia Perez et al. 2016} Garc{\'{\i}}a P{\'e}rez, A.~E., Allende Prieto, C., Holtzman, J.~A., et al.\ 2016, \aj, 151, 144 
\bibitem[Geisler et al.(2012)]{Geisler et al. 2012} Geisler, D., Villanova, S., Carraro, G., et al.\ 2012, \apjl, 756, L40
\bibitem[Gilmore \& Reid(1983)]{Gilmore&Reid 1983} Gilmore, G., \& Reid, N.\ 1983, \mnras, 202, 1025
\bibitem[Gilmore (1984)]{Gilmore 1984} Gilmore, G.\ 1984, \mnras, 207, 223
\bibitem[Gilmore et al.(1985)]{Gilmore et al. 1985} Gilmore, G., Reid, N., \& Hewett, P.\ 1985, \mnras, 213, 257
\bibitem[Gilmore et al.(1989)]{Gilmore et al. 1989} Gilmore, G., Wyse, R. F. G., \& Kuijken, K. 1989, ARA\&A, 27, 555
\bibitem[Gratton et al. (1996)]{Gratton et al. 1996} Gratton, R., Carretta, E., Matteucci, F., \& Sneden, C.\ 1996, Formation of the Galactic Halo...Inside and Out, 92, 307 
\bibitem[Grundahl et al.(2008)]{Grundahl et al. 2008} Grundahl, F., Clausen, J.~V., Hardis, S., \& Frandsen, S.\ 2008, \aap, 492, 171
%\bibitem[Gunn et al.(2006)]{Gunn et al. 2006} Gunn, J.E., et al. 2006, AJ, 131, 2332
\bibitem[Gunn et al.(2006)]{Gunn et al. 2006} Gunn, J.~E., Siegmund, W.~A., Mannery, E.~J., et al.\ 2006, \aj, 131, 2332
\bibitem[Hayden et al.(2015)]{Hayden et al. 2015} Hayden, M.~R., Bovy, J., Holtzman, J.~A., et al.\ 2015, \apj, 808, 132
\bibitem[Hayes et al.(2017)]{Hayes et al. 2017 in prep} Hayes, C.~R. et al., 2017 in prep
\bibitem[Haywood et al.(2013)]{Haywood et al. 2013} Haywood, M., Di Matteo, P., Lehnert, M.~D., Katz, D., \& G{\'o}mez, A.\ 2013, \aap, 560, A109
\bibitem[Holtzman et al.(2015)]{Holtzman et al. 2015} Holtzman, J.~A., Shetrone, M., Johnson, J.~A., et al.\ 2015, \aj, 150, 148 
\bibitem[Jacobson et al.(2011)]{Jacobson et al. 2011} Jacobson, H.~R., Friel, E.~D., \& Pilachowski, C.~A.\ 2011, \aj, 141, 58 
\bibitem[Jacobson et al.(2016)]{Jacobson et al. 2016} Jacobson, H.~R., Friel, E.~D., J{\'{\i}}lkov{\'a}, L., et al.\ 2016, \aap, 591, A37 
\bibitem[Janes \& Phelps (1994)]{Janes&Phelps 1994} Janes, K.A. \& Phelps. R. L. 1994, AJ, 108, 1773
\bibitem[Janes(1979)]{Janes 1979} Janes, K.~A.\ 1979, \apjs, 39, 135
\bibitem[Jilkova et al.(2012)]{Jilkova et al. 2012} Jilkova, L., Carraro, G., Jungwiert, B., \& Minchev, I., 2012, A\&A, 541, A64
\bibitem[Kalirai et al.(2001)]{Kalirai et al. 2001} Kalirai, J.~S., Richer, H.~B., Fahlman, G.~G., et al.\ 2001, \aj, 122, 266
\bibitem[Larson(1976)]{Larson 1976} Larson, R.~B.\ 1976, \mnras, 176, 31
\bibitem[Lee et al.(2015)]{Lee et al. 2015} Lee, D.~M., Johnston, K.~V., Sen, B., \& Jessop, W.\ 2015, \apj, 802, 48 
\bibitem[Leroy et al.(2008)]{Leroy et al. 2008} Leroy, A. K., Walter, F., Brinks, E., et al. 2008, AJ, 136, 2782
\bibitem[Majewski(1993)]{Majewski 1993} Majewski, S.~R.\ 1993, \araa, 31, 575
\bibitem[Majewski et al.(2015)]{Majewski et al. 2015} Majewski, S. R., Schiavon, R. P., Frinchaboy, P. M., et al. 2015, arXiv:1509.05420
\bibitem[Martig et al.(2016)]{Martig et al. 2016} Martig, M., Minchev, I., Ness, M., Fouesneau, M., \& Rix, H.-W.\ 2016, \apj, 831, 139 
%\bibitem[Martig et al.(2016)]{Martig et al. 2016b} Martig, M., Fouesneau, M., Rix, H.-W., et al.\ 2016, \mnras, 456, 3655
\bibitem[Masana et al.(2006)]{Masana et al. 2006} Masana, E., Jordi, C., \& Ribas, I. 2006, A\&A, 450, 735
\bibitem[Masseron \& Gilmore(2015)]{Masseron&Gilmore 2015} Masseron, T., \& Gilmore, G.\ 2015, \mnras, 453, 1855
\bibitem[Matteucci \& Francois (1989)]{Matteucci&Francois 1989} Matteucci, F., \& Francois, P. 1989, MNRAS, 239, 885
\bibitem[Matteucci (2014)]{Matteucci 2014} Matteucci, F.\ 2014, The Origin of the Galaxy and Local Group, Saas-Fee Advanced Course, 37, 145 
\bibitem[Melendez et al.(2008)]{Melendez et al. 2008} Mel{\'e}ndez, J., Asplund, M., Alves-Brito, A., et al.\ 2008, \aap, 484, L21 
\bibitem[Meszaros et al.(2013)]{Meszaros et al. 2013} Meszaros, Sz., Holtzman, J., Garcia Perez, A. E. et al 2013, AJ, 146, 133
%\bibitem[Meszaros et al.(2015)]{Meszaros et al. 2015} Meszaros, Sz. et al., 2015, AJ, 149, 153
\bibitem[M{\'e}sz{\'a}ros et al.(2015)]{Meszaros et al. 2015} M{\'e}sz{\'a}ros, S., Martell, S.~L., Shetrone, M., et al.\ 2015, \aj, 149, 153
\bibitem[Ness et al.(2016)]{Ness et al. 2016} Ness, M., Hogg, D.~W., Rix, H.-W., et al.\ 2016, \apj, 823, 114
%\bibitem[Nidever et al.(2014)]{Nidever et al. 2014} Nidever, D. L., et al., 2014, ApJ, 796, 38
\bibitem[Nidever et al.(2014)]{Nidever et al. 2014} Nidever, D.~L., Bovy, J., Bird, J.~C., et al.\ 2014, \apj, 796, 38
\bibitem[O'Connell(1957)]{O'Connell 1957} O'Connell, D. J. K. 1957, {\it "Stellar Populations"}, RA, 5
\bibitem[Panagia \& Tosi(1980)]{Panagia&Tosi 1980} Panagia, N., \& Tosi, M.\ 1980, \aap, 81, 375
\bibitem[Piskunov et al.(2008)]{Piskunov et al. 2008} Piskunov, A.E., Kharchenko, N.V., Schilbach, E., et al. 2008, A\&A, 487, 557
\bibitem[Platais et al.(2011)]{Platais et al. 2011} Platais, I., Cudworth, K.~M., Kozhurina-Platais, V., et al. 2011, ApJL, 733, L1
\bibitem[Prantzos \& Aubert (1995)]{Prantzos&Aubert 1995} Prantzos, N., \& Aubert, O. 1995, A\&A, 302, 69
\bibitem[Prochaska et al.(2000)]{Prochaska et al. 2000} Prochaska, J.~X., Naumov, S.~O., Carney, B.~W., McWilliam, A., \& Wolfe, A.~M.\ 2000, \aj, 120, 2513
\bibitem[Quinn et al.(1993)]{Quinn et al. 1993} Quinn, P.~J., Hernquist, L., \& Fullagar, D.~P.\ 1993, \apj, 403, 74
\bibitem[Reddy et al.(2006)]{Reddy et al. 2006} Reddy, B.~E., Lambert, D.~L., \& Allende Prieto, C.\ 2006, \mnras, 367, 1329 
\bibitem[Robin et al.(2014)]{Robin et al. 2014} Robin, A.~C., Reyl{\'e}, C., Fliri, J., et al.\ 2014, \aap, 569, A13 
\bibitem[Salaris et al.(2004)]{Salaris et al. 2004} Salaris, M., Weiss, A., \& Percival, S.~M.\ 2004, \aap, 414, 163
\bibitem[Santiago et al.(2016)]{Santiago et al. 2016} Santiago, B.~X., Brauer, D.~E., Anders, F., et al.\ 2016, \aap, 585, A42
\bibitem[Schiavon et al.(2004)]{Schiavon et al. 2004} Schiavon, R.~P., Caldwell, N., \& Rose, J.~A.\ 2004, \aj, 127, 1513
\bibitem[Schultheis et al.(2017)]{Schultheis et al. 2017} Schultheis, M., Rojas-Arriagada, A., Garc{\'{\i}}a P{\'e}rez, A.~E., et al.\ 2017, arXiv:1702.01547 
\bibitem[Sellwood \& Binney(2002)]{Sellwood&Binney 2002} Sellwood, J.~A., \& Binney, J.~J.\ 2002, \mnras, 336, 785
\bibitem[Shetrone et al.(2015)]{Shetrone et al. 2015} Shetrone, M., Bizyaev, D., Lawler, J.~E., et al.\ 2015, \apjs, 221, 24 
\bibitem[Smiljanic et al.(2016)]{Smiljanic et al. 2016} Smiljanic, R., Romano, D., Bragaglia, A., et al.\ 2016, \aap, 589, A115 
\bibitem[Spitzer \& Schwarzschild(1951)]{Spitzer&Schwarzschild 1951} Spitzer, L., Jr., \& Schwarzschild, M.\ 1951, \apj, 114, 385
\bibitem[Tolstoy et al.(2009)]{Tolstoy et al. 2009} Tolstoy, E., Hill, V., \& Tosi, M.\ 2009, \araa, 47, 371 
\bibitem[Wang et al.(2015)]{Wang et al. 2015} Wang, J., Ma, J., Wu, Z., et al. 2015, AJ, 150, 61
\bibitem[Weisz et al.(2014)]{Weisz et al. 2014} Weisz, D.~R., Dolphin, A.~E., Skillman, E.~D., et al.\ 2014, \apj, 789, 147
\bibitem[Wu et al.(2007)]{Wu et al. 2007} Wu, Z.-Y., Zhou, X., Ma, J., et al.\ 2007, \aj, 133, 2061
%\bibitem[Wu et al.(2009)]{Wu et al. 2009} Wu, Zhen-Yu, et al., 2009, MNRAS, 399, 2146–2164
\bibitem[Wu et al.(2009)]{Wu et al. 2009} Wu, Z.-Y., Zhou, X., Ma, J., \& Du, C.-H.\ 2009, \mnras, 399, 2146
\bibitem[Xiang et al.(2017)]{Xiang et al. 2017} Xiang, M.-S., Liu, X.-W., Shi, J.-R., et al.\ 2017, \mnras, 464, 3657 
\bibitem[Yang et al.(2013)]{Yang et al. 2013} Yang, S.-C., Sarajedini, A., Deliyannis, C.P., et al. 2013, ApJ, 762, 3
\bibitem[Yoshii(1982)]{Yoshii 1982} Yoshii, Y.\ 1982, \pasj, 34, 365
\bibitem[Zasowski et al.(2013)]{Zasowski et al. 2013} Zasowski, G., Johnson, J. A., Frinchaboy, P. M. et al. 2013, AJ, 146, 81
\end{thebibliography}
\end{document}